\documentclass[acmsmall,screen=true,authorversion]{acmart}
\pdfoutput=1

\newif\ifextended
\extendedtrue

\AtBeginDocument{%
  \providecommand\BibTeX{{%
    \normalfont B\kern-0.5em{\scshape i\kern-0.25em b}\kern-0.8em\TeX}}}

\setcopyright{rightsretained}
\acmPrice{}
\acmDOI{10.1145/3498665}
\acmYear{2022}
\copyrightyear{2022}
\acmSubmissionID{popl22main-p27-p}
\acmJournal{PACMPL}
\acmVolume{6}
\acmNumber{POPL}
\acmArticle{4}
\acmMonth{1}



\usepackage{math-setup}

\usepackage[mode=build]{standalone}
\usepackage{xspace}
\usepackage{comment}

\usepackage{xparse}

\usepackage[noend]{algpseudocode}

\usepackage{csvsimple}
\usepackage{longtable}
\usepackage{multirow}
\usepackage[nomessages]{fp}
\usepackage{tikz}
\usepackage{cleveref}

\usepackage{threeparttable}
\usepackage{proof}
\usepackage{enumitem}
\usepackage{ulem}
\newcommand{\toolname}{{\sc Solid}\xspace}

\newcommand{\minisol}{{\sc MiniSol}\xspace}

\newcommand{\verismart}{{\sc Verismart}\xspace}

\newcommand{\verismartdata}{{\sc Verismart}\xspace}

\newcommand{\etherscandata}{{\sc Etherscan}\xspace}

\newcommand{\soltype}{{\sc SolType}\xspace}
\newcommand{\etherscan}{{\sc Etherscan}\xspace}
\newcommand{\safemath}{{\sc SafeMath}\xspace}

\newcommand{\toolauto}{{\sc Auto}-\toolname}
\newcommand{\toolsemi}{{\sc Semi}-\toolname}

\newcommand{\vsFpCount}{195} 
\newcommand{\asFpCount}{67} 
\newcommand{\ssFpCount}{46} 
\newcommand{\vsTpCount}{200} 
\newcommand{\asTpCount}{204} 
\FPeval{\vsFpr}{round(100*\vsFpCount/(\vsFpCount+\vsTpCount) :1)} 
\FPeval{\asFpr}{round(100*\asFpCount/(\asFpCount+\asTpCount) :1)} 
\FPeval{\ssFpr}{round(100*\ssFpCount/(\ssFpCount+\asTpCount) :1)}
\newcommand{\totOps}{667}

\newcommand{\EsContractCount}{60}
\newcommand{\EsRedundant}{463}
\newcommand{\vsSafe}{272}
\FPeval{\vsSafePct}{round(100*\vsSafe/\EsRedundant :1)} 
\newcommand{\asSafe}{396}
\FPeval{\asSafePct}{round(100*\asSafe/\EsRedundant :1)} 
\newcommand{\ssSafe}{417}
\FPeval{\ssSafePct}{round(100*\ssSafe/\EsRedundant :1)} 
\newcommand{\zthreeTimeout}{10}

\newcommand{\lineCntAvg}{389}

\newcommand{\vsAvgTime}{425}

\newcommand{\asAvgTime}{24}

\newcommand{\ssAvgTime}{10}

\newcommand{\VsContractCount}{60}
\newcommand{\VsEvalOps}{642}

\newcommand{\VsEvalGtNeg}{390}
\newcommand{\VsEvalVsSafe}{294}
\FPeval{\VsEvalVsSafePct}{round(100*\VsEvalVsSafe/\VsEvalGtNeg :1)}
\newcommand{\VsEvalVsFpr}{27.9}
\newcommand{\VsEvalVsFpCount}{97}
\newcommand{\VsEvalVsTpCount}{251}
\newcommand{\VsEvalVsAvgTime}{476}

\newcommand{\VsEvalAsSafe}{340}
\FPeval{\VsEvalAsSafePct}{round(100*\VsEvalAsSafe/\VsEvalGtNeg :1)}
\newcommand{\VsEvalAsFpr}{16.2}
\newcommand{\VsEvalAsFpCount}{50}
\newcommand{\VsEvalAsTpCount}{252}
\newcommand{\VsEvalAsAvgTime}{62}

\newcommand{\VsEvalSsSafe}{351}
\FPeval{\VsEvalSsSafePct}{round(100*\VsEvalSsSafe/\VsEvalGtNeg :1)}
\newcommand{\VsEvalSsAvgTime}{9}

\newcommand{\VsEvalSsFpCount}{41}
\FPeval{\VsEvalSsFpr}{round(100*\VsEvalSsFpCount / (\VsEvalSsFpCount + \VsEvalAsTpCount) :1)}

\FPeval{\VsEvalSafePctDelta}{round(\VsEvalAsSafePct - \VsEvalVsSafePct :0)}
\FPeval{\EsEvalSafePctDelta}{round(\asSafePct - \vsSafePct :0)}

\FPeval{\TotTotOps}{trunc(\totOps + \VsEvalOps :0)}
\FPeval{\TotTotGtNeg}{trunc(\EsRedundant + \VsEvalGtNeg :0)}
\FPeval{\TotContractCount}{trunc(\EsContractCount + \VsContractCount :0)}

\FPeval{\TotVsSafe}{trunc(\vsSafe + \VsEvalVsSafe :0)}
\FPeval{\TotVsSafePct}{round(100*(\TotVsSafe / \TotTotGtNeg) :1)}
\FPeval{\TotVsFpCount}{trunc(\vsFpCount + \VsEvalVsFpCount :0)}
\FPeval{\TotVsTpCount}{trunc(\vsTpCount + \VsEvalVsTpCount :0)}
\FPeval{\TotVsFpr}{round(100*\TotVsFpCount / (\TotVsFpCount + \TotVsTpCount) :1)}
\newcommand{\TotVsAvgTime}{451}

\FPeval{\TotAsSafe}{trunc(\asSafe + \VsEvalAsSafe :0)}
\FPeval{\TotAsSafePct}{round(100*(\TotAsSafe / \TotTotGtNeg) :1)}
\FPeval{\TotAsFpCount}{trunc(\asFpCount + \VsEvalAsFpCount :0)}
\FPeval{\TotAsTpCount}{trunc(\asTpCount + \VsEvalAsTpCount :0)}
\FPeval{\TotAsFpr}{round(100*\TotAsFpCount / (\TotAsFpCount + \TotAsTpCount) :1)}
\newcommand{\TotAsAvgTime}{41}

\FPeval{\TotSsSafe}{trunc(\ssSafe + \VsEvalSsSafe :0)}
\FPeval{\TotSsSafePct}{round(100*(\TotSsSafe / \TotTotGtNeg) :1)}
\FPeval{\TotSsFpCount}{trunc(\ssFpCount + \VsEvalSsFpCount :0)}
\newcommand{\TotSsTpCount}{\TotAsTpCount}
\FPeval{\TotSsFpr}{round(100*\TotSsFpCount / (\TotSsFpCount + \TotSsTpCount) :1)}

\newcommand{\TotSsAvgTime}{10}

\FPeval{\TotSsSafePctRounded}{round(\TotSsSafePct :0)}

\newcommand{\AblNoInferFpr}{48.7}
\newcommand{\AblNoSoftFpr}{69.4}

\newcommand{\AblNestedFpr}{23.5}
\newcommand{\AblNoNestedFpr}{39.0}

\begin{document}

\title{SolType: Refinement Types for Arithmetic Overflow in Solidity}


\author{Bryan Tan}
\authornote{This author became affiliated with Amazon Web Services during the publication process; however, the work in this paper was performed prior to joining Amazon.}
\affiliation{
  \institution{University of California, Santa Barbara}
  \country{USA}
}
\email{bryantan@cs.ucsb.edu}

\author{Benjamin Mariano}
\affiliation{
  \institution{University of Texas at Austin}
  \country{USA}
}
\email{bmariano@cs.utexas.edu}

\author{Shuvendu K. Lahiri}
\affiliation{
  \institution{Microsoft Research}
  \country{USA}
}
\email{Shuvendu.Lahiri@microsoft.com}

\author{Isil Dillig}
\affiliation{
  \institution{University of Texas at Austin}
  \country{USA}
}
\email{isil@cs.utexas.edu}

\author{Yu Feng}
\affiliation{
  \institution{University of California, Santa Barbara}
  \country{USA}
}
\email{yufeng@cs.ucsb.edu}

\begin{abstract}

As smart contracts gain  adoption in  financial transactions, it becomes increasingly important to ensure that they are free of bugs and security vulnerabilities. Of particular relevance in this context are arithmetic overflow bugs, as integers are often used to represent financial assets like account balances. Motivated by this observation, this paper presents \soltype, a refinement type system for Solidity that can be used to prevent arithmetic over- and under-flows in smart contracts. \soltype allows developers to add refinement type annotations and uses them to prove that arithmetic operations do not lead to over- and under-flows. \soltype incorporates a rich vocabulary of refinement terms that allow expressing relationships between integer values and aggregate properties of  complex data structures.
Furthermore, our implementation, called \toolname, incorporates a type inference engine 
and can automatically infer useful type annotations, including non-trivial contract invariants.

To evaluate the usefulness of our type system, we use \toolname to prove arithmetic safety of a total of {\TotContractCount} smart contracts. When used in its fully automated mode (i.e., using \toolname's type inference capabilities), \toolname is able to eliminate \TotAsSafePct\% of redundant runtime checks used to guard against overflows. We also compare \toolname against a state-of-the-art arithmetic safety verifier called {\sc VeriSmart} and show that \toolname has a significantly lower false positive rate, while being significantly faster in terms of verification time.
\end{abstract}

\maketitle

\section{Introduction}~\label{sec:intro}
Smart contracts are programs that run on top of the blockchain and perform  financial transactions in a distributed environment
without  intervention from trusted third parties. In recent years, smart contracts have seen widespread adoption, with over 45
million~\cite{etherscan} instances covering financial products, online gaming,
real estate~\cite{case1}, shipping, and logistics~\cite{case2}.

Because  smart contracts deployed on a blockchain are freely accessible
through their public methods, any functional bugs or vulnerabilities inside the
contracts can lead to disastrous losses, as demonstrated by recent
attacks~\cite{attack1,attack2,attack3,attack4}. Therefore, unsafe
smart contracts are increasingly becoming a
serious threat to the success of the blockchain technology. For
example, recent infamous attacks on the Ethereum blockchain
such as the DAO~\cite{attack1} and the Parity Wallet~\cite{attack2} attacks were
caused by unsafe smart contracts. To make things worse, smart contracts are
immutable once deployed, so bugs cannot be easily fixed.

One common type of security vulnerability involving smart contracts is arithmetic overflows.
In fact, according to a recent study, such bugs account for over 96\% of CVEs assigned to Ethereum smart contracts~\cite{verismart2020}. Furthermore,
because smart contracts often use integers to represent financial assets, arithmetic bugs, if exploited, can cause significant financial damage~\cite{attack3}. For this reason,  smart contracts rarely \emph{directly} perform arithmetic operations (e.g., addition) but instead perform arithmetic indirectly by calling  a library called \safemath. Since this library inserts run-time checks for over- and under-flows, this approach is effective at preventing exploitable vulnerabilities but comes with  run-time overhead. Furthermore, because computation on the blockchain costs money (measured in a unit called \emph{gas}), using the \safemath library can  significantly increase the cost of running these smart contracts.

Motivated by this observation, we are  interested in \emph{static} techniques that can be used to prevent integer overflows. Specifically,  we propose a refinement type system for Solidity, called \soltype, that can be used to prove that arithmetic operations do not over- or under-flow.  Arithmetic operations that are proven type-safe by \soltype are guaranteed to not overflow; thus, \soltype can be used to eliminate unnecessary runtime checks.




While \soltype is inspired by prior work on logically qualified types~\cite{liquidtype08,liquidtype10,rsc2016}, it needs to address a key challenge that arises in Solidity programs. Because  smart contracts often use non-trivial data structures (e.g., nested mappings or mappings of structs) to store information about accounts, it is important to reason about the relationship between integer values and aggregate properties of such data structures. 
Based on this observation, \soltype allows relational-algebra-like refinements that can be used to perform projections and aggregations over mappings.  While this design choice makes \toolname expressive enough to discharge many potential overflows, it nonetheless enjoys decidable type checking. Furthermore, our implementation, called \toolname, incorporates a type inference engine based on Constrained Horn Clause (CHC) solvers \cite{chc} and can therefore automatically infer useful type annotations, including non-trivial contract invariants.

We evaluate \toolname on 120  smart contracts and compare it against \verismart~\cite{verismart2020}, a state-of-the-art verifier for checking arithmetic safety in smart contracts.
Our evaluation shows that (1) \toolname can discharge \TotAsSafePct\%  of unnecessary runtime overflow checks, and (2) \toolname is both faster and has a lower false positive rate compared to \verismart. We also perform ablation studies to justify our key design choices and empirically validate that they are important for achieving good results.

In summary, this paper makes the following contributions:\looseness=-1
\begin{itemize}[leftmargin=*]
\item We present a new refinement type system called \soltype for proving arithmetic safety properties of smart contracts written in Solidity. Notably, our type system allows expressing relationships between integers and aggregate properties of complex data structures.
\item We implement the proposed type system in a tool called \toolname, which also incorporates a type inference procedure based on Constrained Horn Clause solvers.
\item We evaluate \toolname on 120 real-world smart contracts and show that it is useful for eliminating many run-time checks for arithmetic safety. We also compare \toolname against \verismart and demonstrate its advantages in terms of false positive rate and running time.
\end{itemize}

\section{Overview}~\label{sec:overview}
In the section, we go through the high-level idea of our refinement type system using a representative example.
\subsection{Verifying Overflow Safety: A Simple Example}
Fig.~\ref{fig:erc20-example} shows a snippet from a typical implementation of an ERC20 token. Here,
the \lstinline{mint} function allows an authorized "owner" address to generate tokens, and the \lstinline{transfer} function allows users of the contract to transfer tokens between each other.
User token balances are tracked in the mapping \lstinline{bals}, with the total number of issued tokens  stored in \lstinline{tot}.
It is crucial that the arithmetic does not overflow in either function; otherwise it would be possible for a malicious agent to arbitrarily create or destroy tokens.

This contract does not contain any arithmetic overflows because it maintains the invariant that the summation over the entries in \lstinline{bals} is at most \lstinline{tot}. This invariant, together with the \lstinline{require} clauses in \lstinline{mint} and \lstinline{transfer}, ensures that there are no overflows.  Such \emph{contract invariants} can be expressed as refinement type annotations in \toolname. In particular, as shown in line 4, \lstinline{bals} has the  refinement type $\{ \nu \ | \ \mathrm{sum}(\nu) \leq \mathrm{tot} \}$, which is sufficient to prove the absence of overflows in this program. In the remainder of this section, we illustrate some of the features of \toolname using this example and its variants.

\begin{figure}
\centering
\begin{tabular}{c}
\begin{lstlisting}[language=solidity,escapechar=|]
contract ExampleToken is IERC20 {
    address owner;
    uint tot;
    mapping(address => uint) /* { sum(v) <= tot } */ bals; |\label{line:ex_annot}|
    constructor(address _owner) public {
        owner = _owner;
    }
    
    function mint(uint amt) public {
    	require(msg.sender == owner);
    	require(tot + amt >= tot); |\label{line:ex_mint_check}|
    	tot = tot + amt; |\label{line:ex_mint_add1}|
    	bals[msg.sender] = bals[msg.sender] + amt; |\label{line:ex_mint_add2}|
    }
    
    function transfer(address recipient, uint amt) public {
    	require(bals[msg.sender] >= amt); |\label{line:ex_tran_check}|
    	bals[msg.sender] = bals[msg.sender] - amt; |\label{line:ex_tran_sub}|
    	bals[recipient] = bals[recipient] + amt; |\label{line:ex_tran_add}|
    }
    
    /* ... other functions ... */
}
\end{lstlisting}
\end{tabular}
\caption{Excerpt from a typical ERC20 token.}
\label{fig:erc20-example}
\end{figure}

\subsection{Intermediate Representation}

Due to imperative features like loops and mutation, Solidity is not directly amenable to refinement type checking.
Thus, similar to prior work~\cite{liquidtype10, rsc2016}, we first translate the contract to a intermediate representation (IR) in SSA form that we call MiniSol. Our IR has the following salient features: 
\begin{itemize}[leftmargin=*]
\item Variables are assigned exactly once (i.e., SSA form).
\item Integers are represented as unbounded integers (i.e., integers in $\mathbb{Z}$) rather than as machine integers (i.e., integers modulo $2^{256}-1$).
\item Mappings, structs, and arrays are converted into immutable data structures.
\item A \textit{fetch} statement is used to load the values of the storage variables at function entry, and a \textit{commit} statement is used to store the values of the storage variables at function exit.
\end{itemize}
The resulting IR of the example contract is shown in Fig.~\ref{fig:erc20-example-ir}, which we will use for the examples in this section. {Our tool automatically converts Solidity programs to this IR in a pre-processing step  and automatically inserts \emph{fetch}/\emph{commit} statements at the necessary places. Similar to the \emph{fold}/\emph{unfold} operations from prior work~\cite{liquidtype10}, the use of these \emph{fetch}/\emph{commit} statements in the MiniSol IR  allows temporary invariant violations.}

\begin{figure}
\centering
\begin{tabular}{c}
\begin{lstlisting}[language=minisol,escapechar=|]
contract ExampleToken {
    owner : addr;
    bals : map(addr => uint) /* { sum(v) <= tot } */;
    tot : uint;
    
    constructor(_owner : addr) public {
        let owner0 : addr = _owner;
        let bals0 : map(addr => uint) = zero_map[addr, uint];
        let tot0 : uint = 0;
        commit owner0, bals0, tot0;
    }
    
    fun mint(amt : uint) {
        fetch owner1, bals1, tot1;
    	require(msg.sender == owner1);
    	require(tot1 + amt >= tot1); |\label{line:mint_tot_check}|
    	let tot2 : uint = tot1 + amt; |\label{line:mint_tot_add}|
    	let bals2 : map(addr => uint) =
    	  bals1[msg.sender |$\triangleleft$| bals1[msg.sender] + amt]; |\label{line:mint_bals_add}|
    	commit owner1, bals2, tot2;
    }
    
    fun transfer(recipient : address, amt : uint) {
        fetch owner3, bals3, tot3;
    	require(bals3[msg.sender] >= amt); |\label{line:trans_req}|
    	let bals4 : map(addr => uint) =
    	  bals3[msg.sender |$\triangleleft$| bals3[msg.sender] - amt]; |\label{line:trans_sub}|
    	let bals5 : map(addr => uint) =
    	  bals4[recipient |$\triangleleft$| bals4[recipient] + amt]; |\label{line:trans_add}|
    	commit owner3, bals5, tot3;
    }
}
\end{lstlisting}
\end{tabular}
\caption{MiniSol version of the example token (with some irrelevant details omitted)}
\label{fig:erc20-example-ir}
\end{figure}

\subsection{Discharging Overflow Safety on Local Variables}

We now demonstrate how our refinement type system can be used to verify arithmetic safety  in this example.
Consider the \lstinline{mint} function in Fig.~\ref{fig:erc20-example-ir}.
The require statement on line~\ref{line:mint_tot_check} is a runtime overflow check for \lstinline{tot1 + amt}.
If the runtime check fails, then the contract will abort the current transactions, meaning that it discards any changes to state variables. 
However, \lstinline{tot1 + amt >= tot1} only works as a runtime overflow check when the integers are machine integers.
Since Solid operates on mathematical integers, \toolname will heuristically rewrite common overflow check patterns over machine integers\footnote{Specifically, the patterns include \lstinline{X + Y >= X} and \lstinline{X + Y >= Y}, with similar patterns for multiplication.}, such as the one above, into a predicate over mathematical integers:
\[
    tot1 + amt \leq \pmaxint
\]
where $\pmaxint$ is a constant equal to the maximum machine integer ($= 2^{256}-1$ for Solidity's \lstinline{uint} type).
Since \toolname keeps track of this predicate as a \textit{guard predicate}, it is able to prove the safety of the addition on line~\ref{line:mint_tot_add}.

Next,  we explain how \toolname proves the safety of the addition  on line~\ref{line:mint_bals_add}. For now, let us assume that the refinement type annotation for \lstinline{bals} is indeed valid, i.e., \lstinline{bals1} has the following refinement type (we omit base type annotations for easier readability):
\begin{equation}
    bals1 : \rtypei{\nu \ | \ \psum(\nu) \leq tot1}
    \label{eq:ex_bals1_type}
\end{equation}
Since \lstinline{bals1} contains only unsigned integers, we refine \lstinline{bals1[msg.sender]} with the type:
$$bals1[msg.sender] : \rtypei{\nu \ | \ \nu \leq \psum(bals1)}$$
From transitivity of $\leq$, this implies that  $bals1[msg.sender] \leq tot1$. Finally, using the guard predicate $tot1+amt \leq \pmaxint$, \toolname is able to prove:
$$bals1[msg.sender] + amt \leq \pmaxint$$
which implies that the addition does not overflow. 
In a similar manner, \toolname can also prove the safety of the operations in \lstinline{transfer} using the refinement type of state variable \lstinline{bals} and the \lstinline{require} statement at line 25.

\subsection{Type Checking the Contract Invariant}
\label{sec:overview_cinv_ex}
In the previous subsection, we showed that contract invariants (expressed as refinement type annotations on state variables) are useful for discharging overflows. However, \toolname needs to establish that these type annotations are valid, meaning that contract invariants are preserved at the end of functions.  For instance, at the end of the \lstinline{mint} function, the type checker needs to establish that \lstinline{bals2} has the following refinement type:
\begin{equation}
bals2 : \rtypei{\nu \ | \ \psum(\nu) \leq tot2} \label{eq:ex_check_inv}
\end{equation}

To establish this, we proceed as follows: First, from line \ref{line:mint_bals_add}, \toolname infers the type of \lstinline{bals2} to be:
\begin{equation}
\begin{split}
bals2 : \rtypei{\nu \ | 
\psum(\nu) = \psum(bals1) + amt)
}
\end{split}
\label{eq:ex_bals2_type}
\end{equation}
Then, since \lstinline{tot2} has the refinement type $\{ \nu \ |\ \nu = tot1 + amt \}$ and \lstinline{bals1} has the refinement type $\{ v \ | \ \psum(\nu) \leq tot1 \}$, we can show that $\psum(bals2) \leq tot2$. This establishes the desired contract invariant (i.e., Eq.~\ref{eq:ex_check_inv}).  Using similar reasoning, \toolname can also type check that the contract invariant is preserved in  \lstinline{transfer}.

\subsection{Contract Invariant Inference}

Although the type annotation is  given explicitly in this example, \toolname can actually prove the absence of overflows without \textit{any} annotations by performing type inference. To do so, \toolname first introduces a ternary uninterpreted predicate $I$ (i.e., unknown contract invariant) for each state variable (\lstinline{owner, bals, tot}) that relates its value $\nu$ to the other two state variables. This corresponds to the following type "annotations" for the state variables:
\begin{align*}
    owner &: \rtypei{\nu \ | \ I_1(\nu, tot, bals)} \\
    tot &: \rtypei{\nu \ | \ I_2(owner, \nu, bals)} \\
    bals &: \rtypei{\nu \ | \ I_3(owner, tot, \nu)}
\end{align*}

Given these type annotations over the unknown predicates $I_1$-$I_3$, \toolname generates constraints on $I_1$-$I_3$ during type inference and leverages a CHC solver to find an instantiation of each $I_i$. However, one complication is that we cannot feed all of these constraints to a CHC solver because the generated constraints may be unsatisfiable. In particular, consider the scenario where the program contains a hundred arithmetic operations, only one of which is unsafe. If we feed all constraints generated during type checking to the CHC solver, it will correctly determine that there is no instantiation of the invariants that will allow us to prove the safety of \emph{all} arithmetic operations. However, we would still like to infer a type annotation that will allow us to discharge the remaining 99 arithmetic operations.

To deal with this difficulty, \toolname tries to solve the CHC variant of the well-known MaxSAT problem. That is, in the case where the generated Horn clauses are unsatisfiable, we would like to find an instantiation of the unknown predicates so that the maximum number of constraints are satisfied. \toolname solves this "MaxCHC"-like problem by generating one overflow checking constraint at a time and iteratively strengthening the inferred contract invariant. For instance, for our running example,  \toolname can automatically infer the desired contract invariant $I_3$, namely \lstinline{sum(bals) <= tot} ($I_1$ and $I_2$ are simply $\etrue$).

\subsection{Refinements for Aggregations over Nested Data Structures} \label{sec:overview_sum_data_structure}

A unique aspect of our type system is its ability to express more complex relationships between integer values and data structures like nested mappings over structs. To illustrate this aspect of the type system,
consider Fig.~\ref{fig:erc20-example-struct}, which is a modified version of Fig.~\ref{fig:erc20-example} that stores balance values inside a \lstinline{User} struct in a nested mapping.
The addition on line~\ref{line:struct_ex_add} cannot overflow because the contract maintains the invariant that the summation over the \lstinline{bal} fields of all the nested \lstinline{User} entries in \lstinline{usrs} is at most \lstinline{tot}.
\toolname is able to infer that the type of $usrs$ is
\[
\rtypei{\nu \mid \psum(\pfld_{bal}(\pflat(\nu))) \leq tot}
\]

{In particular, this refinement type captures the contract invariant that the sum of all user balances nested inside the struct of \lstinline{usrs}  is bounded by \lstinline{tot}. Using such a refinement type, we can prove the safety of the addition on line~\ref{line:struct_ex_add}. Note that the use of operators like $\mathsf{Flatten}$ and $\mathsf{Fld}_\emph{bal}$ is vital for successfully discharging the potential overflow in this example.}


\begin{figure}
\begin{tabular}{c}
\begin{lstlisting}[language=minisol, escapechar=|]
contract ExampleTokenWithStruct {
    struct User {
        uint bal;
        /* ... other fields ... */
    }
    
    /* ... other state variables ... */
    mapping(address => mapping(uint => User)) usrs;
    
    function mint(uint amt, uint accno) public {
    	require(msg.sender == owner);
    	require(tot + amt >= tot);
    	tot += amt;
    	usrs[msg.sender][accno].bal += amt; |\label{line:struct_ex_add}|
    }
    
    /* ... other functions ... */
}
\end{lstlisting}
\end{tabular}
\caption{The Solidity source code of the running example (Fig.~\ref{fig:erc20-example}) modified to use nested data structures.}
\label{fig:erc20-example-struct}
\end{figure}

\section{Language}~\label{sec:lang}

\begin{figure}[t]
  \include{figures/syntax_lang}
  \caption{Syntax of \minisol programs}
  \label{fig:syntax}
\end{figure}

In this section, we present the syntax of \minisol (Figures~\ref{fig:syntax} and~\ref{fig:syntax_types}), an intermediate representation for modeling Solidity smart contracts. 
In what follows, we give a brief overview of the important features of \minisol relevant to the rest of the paper.

\subsection{\minisol Syntax}

In \minisol, a program is a contract $C$  that contains fields (referred to as \emph{state variables}), struct definitions $S$, and function declarations $f$. State variables are declared and initialized in a constructor $ctor$. Functions include a type signature, which provides a type $\type_i$ for each argument $x_i$ as well as return type $\type_r$, and a method body, which is a sequence of statements, followed by an expression.

Statements in \minisol include let bindings, conditionals, function calls (with pass-by-value semantics), assertions, and assumptions.
As \minisol programs are assumed to be in SSA form, we include a \emph{join point} $j$ for if statements and while loops. In particular, a  join point consists of a list of $\Phi$-node variable declarations, where each $x_i$ is declared to be of type $\type_i$ and may take on the value of either $x_{i1}$ or $x_{i2}$, depending on which branch is taken.
In line with this SSA assumption, we 
also assume variables are not redeclared.

As mentioned previously, another feature of \minisol is that it contains fetch and commit statements for  reading the values of all state variables from the blockchain and writing them to the blockchain respectively.
Specifically, the construct 
\[
\sfetch{x_1' \sas x_1, \dots, x_n' \sas x_n}
\]
\emph{simultaneously} retrieves the values of state variables $x_1', \ldots, x_n'$ and writes them into local variables $x_1, \ldots, x_n$.
Similarly, the commit statement 
\[
\scommit{e_1 \sto x_1, \dots, e_n \sto x_n} 
\]
simultaneously stores the result of evaluating expressions $e_1, \ldots, e_n$ into state variables $x_1, \ldots, x_n$.
Note that Solidity does not contain such fetch and commit constructs; however, we include them in \minisol to allow temporary violations of contract invariants inside procedures.
Such mechanisms have also been used in prior work \cite{liquidtype10,alias-types} for similar reasons.

Expressions  in \minisol consist of variables, unsigned integer and boolean constants, binary operations, and data structure operations.  Data structures include maps and structs, and we use the same notation for accessing/updating structs and maps.
In particular, $e_1[e_2]$ yields the value stored at key (resp. field) $e_1$ of map (resp. struct) $e_2$.
Similarly, $e_1[e_2 \triangleleft e_3]$ denotes the new map (resp. struct) obtained by writing value $e_3$ at key (resp. field) $e_2$ of $e_1$. 
For structs, we use a special type of expression called a \emph{field selector} to indicate struct access (e.g. $e_1[.x]$).

\subsection{Types}

\begin{figure}[t]
  \include{figures/syntax_types}
  \caption{Syntax of \minisol types}
  \label{fig:syntax_types}
\end{figure}

Similar to previous refinement type works, \minisol provides both \emph{base type} annotations as well as \emph{refinement type} annotations (Figure~\ref{fig:syntax_types}). Base types correspond to standard Solidity types, such as unsigned integers (256-bit), booleans, mappings, and structs. We shorten "mappings" to $\mathsf{Map}$ and assume that all keys are $\tint$s, as most  key types in Solidity mappings are coercible to $\tint$.
Thus, $\tmapping{\btype}$ is a mapping from unsigned integers to $\btype$s.

A refinement type $\rtype{T}{\phi}$ refines the base type $T$ with a logical qualifier $\phi$. In this paper, logical qualifiers belong to the  quantifier-free combined theory of rationals, equality with uninterpreted functions,  and arrays. Note that determining validity in this theory is decidable, and we intentionally use the theory of rationals rather than integers to keep type checking decidable in the presence of non-linear multiplication. In more detail, logical qualifiers  are boolean combinations of binary relations $\oplus$ over refinement terms $t$, which include both \minisol expressions as well as terms containing the special constructs   $\psum$, $\pfld$, and $\pflat$ (represented as uninterpreted functions) that operate over terms of type $\mathsf{Map}$. In particular,
\begin{itemize}[leftmargin=*]
    \item $\psum(t)$ represents the sum of all values in $t$.
    \item  Given a mapping $t$ containing structs, $\pfld_x(t)$ returns a mapping that only contains field $x$ of the struct. Thus, $\pfld$ is similar to the projection operator from relational algebra.
\item Given a nested mapping $t$, $\pflat(t)$ flattens the map. In particular, given a nested mapping $\mathsf{Map}(\mathsf{Map}(t))$, $\pflat$ produces a mapping $\mathsf{Map}(t)$ where each value of the mapping corresponds to a value of one of the nested mappings from $\mathsf{Map}(\mathsf{Map}(t))$.
\end{itemize}
    
In addition to these constructs, refinement terms in \minisol also include a constant called $\pmaxint$, which indicates  the largest integer value representable in $\tint$.

\begin{example}
    Consider the \lstinline{usrs} mapping in Fig.~\ref{fig:erc20-example-struct}.
    The predicate
    \[
    \psum(\pfld_{bal}(\pflat(usrs))) \leq \pmaxint
    \]
    expresses that if we take the \lstinline{usrs} mapping (which is of type $\tmapping{\tmapping{\emph{User}}}$) and (a) flatten the nested mapping so that it becomes a non-nested mapping of $\emph{User}$; (b) project out the \lstinline{bal} field of each struct to obtain a $\tmapping{\tint}$; and (c) take the sum of values in this final integer mapping, then the result should be at most $\mathsf{MaxInt}$.
\end{example}

\section{Type system}~\label{sec:type-rules}
In this section, we introduce the \soltype refinement type system for the \minisol language.

\subsection{Preliminaries}

We start by discussing the environments and judgments used in our type system; these are summarized in Fig.~\ref{fig:env}.

\begin{figure}
\include{figures/syntax_type_judg}
\caption{Environments and judgments used in our typing rules}
\label{fig:env}
\end{figure}

\paragraph{Environments} Our type system employs a local environment $\Env$ and a set of \emph{guard} predicates $\Guards$.
A local environment $\Env$ is a sequence of type bindings (i.e., $x : \tau$), and
we use the notation $\dom(\Env)$ to denote the set of variables bound in $\Env$.
The guard predicates $\Guards$ consist of (1) a set of path conditions under which an expression is evaluated and (2) a \emph{lock status} $L$ ($\mathsf{Locked}$ or $\mathsf{Unlocked}$) that tracks whether the state variables are currently being modified.

In addition to $\Env$ and $\Guards$, our typing rules need access to struct and function definitions. For this purpose, we make use of another environment $\Delta$ that stores function signatures, struct definitions, and types of the state variables.
To avoid cluttering  notation, we assume that $\Delta$ is implicitly available in the expression and statement typing judgments, as $\Delta$ does not change in a function body.
We use the notation $\Delta(S)$ to indicate looking up the fields of  struct $S$, and we write $\Delta(S, x)$ to retrieve the type of field $x$ of struct $S$.
Similarly, the notation $\Delta(f)$ yields  the function signature of $f$, which consists of  a sequence of argument type bindings as well as the return type.
Finally, given a state variable $x$, $\Delta(x)$ returns the type of $x$.

\paragraph{Typing Judgments}  As shown in Fig.~\ref{fig:env}, there are three different typing judgments in \minisol:

\begin{itemize}[leftmargin=*]
    \item \emph{Expression typing judgments}: We use the judgment $\tjudg{\Env;\Guards}{e}{\tau}$ to type check a \minisol expression $e$. In particular, this judgment indicates that expression $e$ has refinement type $\tau$ under local environment $\Env$, guard $\Guards$, and (implicit) global environment $\Delta$. 
    \item \emph{Statement typing judgments}: Next, we use a judgment of the form 
    $\tjudgS{\Env;\Lock,\Guards}{s}{\Env';\Lock',\Guards'}$ to type check statements. The meaning of this judgment is that statement $s$ type checks under local environment $\Env$, lock status $\Lock$, and guard $\Guards$, and it produces a new local type environment $\Env'$, lock status $\Lock'$, and guard $\Guards'$. Note that the statement typing judgments are flow-sensitive in that they modify the environment, lock status, and guard; this is because (1) let bindings can add new variables to the local type environment, and (2) fetch/commit statements will modify the lock status and guard.
    When the guards are empty, we omit them from the judgment to simplify the notation.
 \item \emph{Join typing judgments}: Finally, we have a third typing judgment for join points of conditionals.
 This judgment is of the form $\tjudgS{\Env;\Guards;(\Env_1;\Guards_1);(\Env_2;\Guards_2)}{j}{\Env'}$ where $\Env, \Guards$ pertain to the state before the conditional, $\Env_i, \Guards_i$ pertain to the state after executing each of the branches, and $\Env'$ is the resulting environment after the join point.
\end{itemize}


\paragraph{Subtyping} Finally, our type system makes use of a subtyping judgment of the form:
$$\subty{\Env;\Guards}{\tau_1}{\tau_2}$$
which states that, under $\Env$ and $\Guards$, the set of values represented by $\tau_1$ is a subset of those represented by $\tau_2$.
As expected and as shown in Fig.~\ref{fig:subtype}, the subtyping judgment reduces to checking logical implication.
In the {\sc Sub-Base} rule, we use a function $\mathsf{Encode}$ to translate $\Env, \gamma$ and the two qualifiers $\phi_1, \phi_2$ to logical formulas which belong to the quantifier-free fragment of the combined theory of rationals, arrays, equality with uninterpreted functions. Thus, to check whether $\tau_1$ is a subtype of $\tau_2$, we can use an off-the-shelf SMT solver.
Since our encoding of refinement types into SMT is the standard scheme used in \cite{liquidtype08,liquidtype10,rsc2016}, we do not explain the $\mathsf{Encode}$ procedure in detail.

\begin{example}
Consider the following subtyping judgment:
\[
\subty{\Gamma; c\geq d}
{\rtype{\tint}{\nu = a}}{\rtype{\tint}{\nu \geq d}}
\]
where $\Gamma$ contains four variables $a,b,c,d$ all with base type $\tint$ and $a$ has the refinement $\nu = b+c$.
This subtyping check reduces to querying the validity of the following formula:
\[
\phi_\btype \land a = b + c \land c \geq d \land \nu = a
\implies
\nu \geq d
\]
where $\phi_\btype$ are additional clauses that restrict the terms of type $\tint$ to be in the interval $[0, \pmaxint]$.
\end{example}

\begin{figure}[t]
    \centering
    \include{figures/judg-subtype}
    \caption{Subtyping relation}~\label{fig:subtype}
\end{figure}

\subsection{Refinement Typing Rules}\label{sec:check-basic}

In this section, we discuss the basic typing rules of \soltype, starting with expressions.
Note that the typing rules we describe in this subsection are \emph{intentionally} imprecise for complex data structures involving nested mappings or mapping of structs.
Since precise handling of complex data structures requires introducing additional machinery, we delay this discussion until the next subsection. 


\subsubsection{Expression Typing}

\begin{figure}
  \include{figures/typing_expr}
  \vspace{-0.1in}
  \caption{Main refinement typing rules for expressions}~  \label{fig:typing_expr}
\vspace{-0.1in}
\end{figure}

Fig.~\ref{fig:typing_expr} shows the key typing rules for expressions. Since the first two rules  and the rules involving structs are standard, we focus on the remaining rules for arithmetic expressions and mappings.

\paragraph{Arithmetic expressions.} The \textsc{TE-Plus} and \textsc{TE-Mul} rules  capture the overflow safety properties of 256-bit integer addition and multiplication respectively. In particular,  they check that the sum/product of the two expressions is less than $\mathsf{MaxInt}$ (when treating them as mathematical integers).  The next rule, \textsc{TE-Minus}, checks that the result of the subtraction is not negative (again, when treated as mathematical integers). Finally, the {\sc TE-Div} rule disallows division by zero, and constrains the result of the division to be in the appropriate range.



\paragraph{Mappings.} The next two rules refer to reading from and writing to mappings. As stated earlier, we only focus on precise typing rules for variables of type $\mathsf{Map(UInt)}$ in this section and defer precise typing rules for more complex maps to the next subsection. Here, according to the  \textsc{Te-MapInd} rule, the result of evaluating $e_1[e_2]$ is less than or equal to $\mathsf{Sum}(e_1)$, where $\mathsf{Sum}$ is an uninterpreted function. According to the next rule, called    \textsc{Te-MapUpd}, if we write value $e_3$ at index $e_2$ of mapping $e_1$, the sum of the elements in the resulting mapping is given by:
\[
\mathsf{Sum}(e_1) - e_1[e_2] + e_3
\]
Thus, this rule, which is based on an axiom from a previous work \cite{verx2019}, allows us to precisely capture the sum of elements in mappings whose value type is $\tint$.

\subsubsection{Statement Typing}

\begin{figure}
  \include{figures/typing_stmt}
  \caption{Main typing rules for statements and function definitions}
  \label{fig:typing_stmt}
\end{figure}

Next, we consider the statement typing rules shown in Fig.~\ref{fig:typing_stmt}. Again, since some of the rules are standard, we focus our discussion on aspects that are unique to our type system.

\paragraph{Conditionals.} In the conditional rule {\sc TS-If}, we separately type check the true and false branches, adding the appropriate predicate (i.e., $e$ or $\neg e$) to the statement guard for each branch. The results of the two branches are combined using the join typing judgment {\sc T-Join} (also shown in Fig.~\ref{fig:typing_stmt}). In particular, the join rule checks the type annotations for variables that are introduced via  $\Phi$ nodes in SSA form and requires that the type of the $x_i$ variants in each branch is a subtype of the declared type $\tau_i$ of $x_i$. Note also that the join rule requires agreement between the lock states in the two branches. This means that either (1) both branches commit their changes to state variables, or (2) both have uncommitted changes.

\paragraph{Fetch and Commit.} As mentioned earlier, \minisol contains $\mathsf{fetch}$ and $\mathsf{commit}$ statements, with the intention of allowing temporary violations of contract invariants. The {\sc TS-Fetch} and {\sc TS-Commit} rules provide type checking rules for these two statements. Specifically, the \textsc{TS-Fetch} rule says that, when no ``lock" is held on the state variables (indicated by a lock status $\unlocked$), a fetch statement can be used to ``acquire" a lock and copy the current values of the state variables into a set of freshly declared variables.
The contract invariant will be assumed to hold on these variables.

Next, according to the \textsc{TS-Commit} rule, a commit statement can  be used to ``release" the lock and write the provided expressions back to the state variables.
The contract invariant will be checked to hold on the provided expressions. Thus, our fetch and commit constructs allow temporary violations of the contract invariant \emph{in between} these fetch and commit statements.

\paragraph{Function Calls.} The {\sc TS-Call} rule in Fig.~\ref{fig:typing_stmt} is used to type check function calls. In particular, since the contract invariant should not be violated between different transactions, our type system enforces that all state variables have been committed by requiring that the current state is $\mathsf{Unlocked}$. Since the type checking of function arguments and return value is standard, the rest of the rule is fairly self-explanatory.


\paragraph{Function Definitions.} Finally, we consider the type checking rule for function definitions (rule \textsc{T-FunDecl} rule in Fig.~\ref{fig:typing_stmt}). According to this rule,
a function declaration is well-typed if its body $s$ is well-typed (with the environment initially set to the arguments and lock status being  $\unlocked$), and the returned expression has type $\type_r$ after executing $s$.

\subsection{Generalizing Sum Axioms to Deeply Nested Mappings} \label{sec:agg-props}

The \textsc{TE-MapInd} and \textsc{TE-MapUpd} rules that we presented in Fig.~\ref{fig:typing_expr} are sufficient for handling maps over integers; however, they are completely imprecise for more complex data structures, such as nested mappings or mappings of structs. Since Solidity programs often employ deeply nested mappings involving structs, it is crucial to have precise typing rules for more complex data structures. Therefore, in this section, we show how to generalize our precise typing rules for mappings over integers to arbitrarily nested mappings.

To gain some intuition about how to generalize these typing rules, consider the \lstinline{usrs} mapping in Fig.~\ref{fig:erc20-example-struct}.
The update to \lstinline{usrs} on line~\ref{line:struct_ex_add} is expressed in our IR as a sequence of map lookup operations followed by a sequence of  updates:
\begin{lstlisting}[language=minisol, escapechar=|, frame=none, numbers=none]
//  usrs0 has type mapping(address => mapping(uint => User))
let m0 = usrs0[msg.sender];         // lookup
let u0 = m0[accno];                 // lookup
let b0 = u0[.bal];                  // access bal field
let u1 = u0[.bal |$\triangleleft$| b0 + amt];       // update bal field of struct
let m1 = m0[accno |$\triangleleft$| u1];            // update usrs[msg.sender] with updated struct
let usrs1 = usrs0[msg.sender |$\triangleleft$| m1]; // update usrs with updated nested mapping
\end{lstlisting}

Before we consider the revised typing rules, let us first consider the constraints we would \emph{want} to generate for this example:

\begin{itemize}[leftmargin=*]
\item For the first line,  \lstinline{usrs0} is a nested mapping containing a struct called \lstinline{User} which has an integer field called \lstinline{bal}, and \lstinline{m0} is a mapping over \lstinline{User}s.  Based on the semantics of our refinement terms, the assignment on the first line imposes the following constraint on  \lstinline{m0}: 
\[
\psum(\pfld_{bal}(m0)) \leq \psum(\pfld_{bal}(\pflat(usrs0)))
\]

\item Using similar reasoning, we can infer the following constraint on \lstinline{u0}, which is of type \lstinline{User}: 
 \[ u0[.bal] \leq \psum(\pfld_{bal}(m0)) \]
\item When we update \lstinline{m1} on the fourth line, this again imposes a constraint involving $\psum$:
    \[
    \begin{split}
       \psum(\pfld_{bal}(m1)) = \psum(\pfld_{bal}(m0)) - m0[accno][.bal] + u1[.bal]
    \end{split}
    \]
\item Finally, for the update on the last line, we can infer:
    \[
    \begin{array}{l}
    \psum(\pfld_{bal}(\pflat(usrs1))) =
    \\ \qquad
    \psum(\pfld_{bal}(\pflat(m0))) - \psum(\pfld_{bal}(m0)) + \psum(\pfld_{bal}(m1))
    \end{array}
    \]
\end{itemize}

As we can see from this example, the \emph{shape} of the constraints we infer for complex map reads and writes is very similar to what we had in the {\sc TE-MapInd} and the {\sc TE-MapUpd} rules from Fig.~\ref{fig:typing_expr}. In particular, reading from a map imposes constraints of the form: 
\begin{equation}
  \hole_2(m[k]) \leq \hole_1(m)
  \label{eq:sctx_lookup_template}
\end{equation}
whereas updating a map imposes constraints of the form:
\begin{equation}
  \hole_1(\eupd{m}{k}{v}) = \hole_1(m) - \hole_2(m[k]) + \hole_2(v)
  \label{eq:sctx_update_template}
\end{equation}
Therefore, we can easily generalize  the {\sc TE-MapInd} and the {\sc TE-MapUpd} rules to arbitrarily complex mappings as long as we have a way of figuring out how to instantiate $H_1, H_2$ in Equations~\ref{eq:sctx_lookup_template} and~\ref{eq:sctx_update_template}.

\subsubsection{Refinement Term Templates}
Based on the above observation, given a source expression $e$ of some base type $T$, we want a way to automatically generate relevant refinement terms of the form $\hole(e)$. To facilitate this, we first introduce the notion of \emph{templatized refinement terms}:

\begin{definition}{\bf (Templatized refinement term)} A \emph{ refinement term template} $\hole$ is a refinement term containing a unique hole, denoted $\square$. Given such a template $\hole$, we write $\hole(e)$ to denote the refinement term obtained by filling the hole $\square$ in $\hole$ with $e$.

\end{definition}

For instance, $\hole = \psum(\pfld_\emph{bal}(\square))$ is a valid template, and $\hole(m0)$ yields $ \psum(\pfld_\emph{bal}(m0))$.

Next, given a type $T_h$, we need a way to generate all templates that can be applied to expressions of type $T_h$. Towards this purpose, we introduce a \emph{template synthesis judgment} of the following form:
\[
\begin{array}{rllr}
    \multicolumn{3}{l}{\tjudgP{\btype_h}{\hole}{\btype}{w}} & \text{(Template synthesis judgment)}
\end{array}
\]
The meaning of this judgment is that, given an expression $e$ of type $T_h$, $\hole(e)$ is a well-typed term of type $T$. In this judgment, $w$ is a so-called \emph{access path} that keeps track of the field accesses we need to perform to  get from $e$ to $\hole(e)$. In particular, an access path is a string over the language of field names, defined according to the following grammar:
\[
\begin{array}{rllr}
   w & ::= & \epsilon \mid xw & \text{(Access path)} \\ \\
\end{array}
\]
where $x$ is the name of a field. As we will see shortly, we need this concept of access path to ensure that we do not generate non-sensical constraints in our generalized type checking rules for map reads and writes. 

\begin{figure}
  \input{figures/typing_paths}
  \caption{Inference rules for type-directed synthesis of refinement term templates}
  \label{fig:hole_rules}
\end{figure}

Fig.~\ref{fig:hole_rules} shows our rules for synthesizing refinement term templates.
The first rule, {\sc TH-Sum}, states that we can apply the $\psum$ operator to an expression whose base type is $\tmapping{\tint}$.
The second rule, {\sc TH-Flatten}, allows applying a $\pflat$ operation to any term of type $\tmapping{\tmapping{\btype}}$.
The third rule, {\sc TH-Fld}, states that we can apply a projection operator to an expression that is a mapping of structs.
In other words, if $\hole$ is a mapping over structs $S$ and $S$ has a field called $x$, then we can generate the template $\pfld_x(\hole)$.
Since this involves a field access, note that the {\sc TH-Fld} rule adds $x$ to the access path.
The next rule, {\sc Th-FldAcc}, is very similar but applies to structs instead of mappings of structs.
In particular, if $\hole$ has struct type $S$ with field $x$, it is valid to access the $x$ field of $H$.
As in the previous case, this rule appends $x$ to the resulting access path.
The final rule, {\sc TH-Hole} is a base case and states that the hole is constrained to have the specified type $\btype_h$.

\subsubsection{Generalized Typing Rules for Map Reads and Writes}

\begin{figure}
    \centering
    \begin{mathpar}
      \input{figures/rule_map_with_holes}
    \end{mathpar}
    \caption{Updated typing rules for mappings, where refinement term templates are used to generate sum properties over nested data structures.}
    \label{fig:new_map_typing_rules}
\end{figure}

Equipped with the template synthesis rules, we are now ready to present the generalized and precise versions of \textsc{TH-MapInd} and \textsc{TH-MapUpd} rules for arbitrarily complex mappings.

The new {\sc TE-MapInd} rule shown in Fig.~\ref{fig:new_map_typing_rules} generates precise constraints on $e_1[e_2]$ where $e_1$ is has type $\mathsf{Map}(T)$. As illustrated earlier through the example, this rule generates constraints of the form $\hole_2(\nu) \leq \hole_1(e_1)$, where $\hole_1, \hole_2$ are  templates that can be applied to terms of type $\mathsf{Map}(T)$ and $T$ respectively. Since a given struct can have multiple integer fields, this rule generates a conjunction of such constraints, one for each integer field. Note that, for each constraint of the form $H_{i2}(\nu) \leq H_{i1}(e_1)$ in this rule, we enforce that the corresponding access paths $w_i$ match, as, otherwise, the generated constraints would not make sense.

Next, we consider the new {\sc TE-MapUpd} rule in Fig.~\ref{fig:new_map_typing_rules} for updating maps. Specifically, given an expression $e_1[e_1 \triangleleft e_3]$ where $e_1$ is of type $\mathsf{Map}(T)$,  this rule generates a conjunction of constraints of the form:
\[
H_{i1}(\nu) = H_{i2}(e_1) - H_{i2}(e_1[e_2]) + H_{i2}(e_3), 
\]
one for each integer field accessible from $T$. As in the previous case, we use the notion of access paths to ensure that the hole templates $H_{i1}$ and $H_{i2}$ correspond to the same field sequence.

\begin{example}
Consider again  the update to \lstinline{usrs} on line~\ref{line:struct_ex_add} of Fig~\ref{fig:erc20-example-struct}.
We demonstrate how the \textsc{TE-MapInd} rule generates typing constraints.
First, since \lstinline{usrs} has type $\tmapping{\tmapping{\emph{User}}}$, we have $\btype = \tmapping{\emph{User}}$.
From the synthesis rules, one such instantiation for $H_{i1}$ is $\psum(\pfld_{bal}(\pflat(\holet)))$, as shown by the following derivation tree:
\begin{center}
\small
\begin{mathpar}
  \inferrule*[Left=TH-Sum]{
    \inferrule*[Left=TH-Fld]{
      \inferrule*[Left=TH-Flatten]{
        \inferrule*[Left=TH-Hole]{ }{
          \tjudgP{\tmapping{\tmapping{{User}}}}{\holet}{\tmapping{\tmapping{{User}}}}{\epsilon}
        }
      } {
        \tjudgP{\tmapping{\tmapping{{User}}}}{\pflat(\holet)}{\tmapping{User}}{bal}
      }
      \\
      \defns(User, bal) = \tint
    } {
    \tjudgP{\tmapping{\tmapping{{User}}}}{\pfld_{bal}(\pflat(\holet))}{\tmapping{\tint}}{bal}
    }
  } {
    \tjudgP{\tmapping{\tmapping{{User}}}}{\psum(\pfld_{bal}(\pflat(\holet)))}{\tint}{bal}
  }
\end{mathpar}
\end{center}
Next, we find an instantiation for $H_{i2}$ that accesses the same fields $w = bal$.
It can be proven that
$$\tjudgP{\tmapping{\tstruct}}{\psum(\pfld_{bal}(\holet))}{\tint}{bal}$$
which yields $H_{i2} = \psum(\pfld_{bal}(\holet))$.
Then, using  the \textsc{TE-MapInd} rule, we obtain the constraint
\begin{equation}
    \psum(\pfld_{bal}(\nu)) \leq \psum(\pfld_{bal}(\pflat(usrs)))
    \label{eq:hole_ex_p1}
\end{equation}
Note that this is only \emph{one} of the predicates that we can derive. For example, if the \lstinline{User} had another integer field called \lstinline{frozen}, we would also generate the following predicate:
\begin{align}
    \psum(\pfld_{frozen}(\nu)) &\leq \psum(\pfld_{frozen}(\pflat(usrs)))
    \label{eq:hole_ex_p2}
\end{align}
Taking the conjunction of Eqs.~\eqref{eq:hole_ex_p1} and \eqref{eq:hole_ex_p2}, we obtain the following type for \lstinline{usrs[msg.sender]}:
\begin{align*}
    usrs[msg.sender] : \rtypei{
        \nu \mid
        & \nu = usrs[msg.sender] \\
        & \land \psum(\pfld_{bal}(\nu)) \leq \psum(\pfld_{bal}(\pflat(usrs))) \\
        & \land \psum(\pfld_{frozen}(\nu)) \leq \psum(\pfld_{frozen}(\pflat(usrs)))
    }
\end{align*}
\end{example}


\subsection{Soundness}

We characterize  the soundness of our type system in the typical way through progress and preservation theorems for both expressions and statements.
We briefly state the key propositions and refer the reader to the extended version of the paper \cite{extended-version} for details.

The execution of a \minisol program can be modeled as evaluation of a closed statement.
Motivated by this observation, expressions are evaluated under the empty environment and empty guard predicates.

\begin{proposition}[Progress for Expressions]
  \label{prop:progress-expr}
  If \ $\tjudg{\PrfCtx}{\expr}{\type}$, then $\expr$ is a value or there exists an $\expr'$ such that $\stepE{\expr}{\expr'}$.
\end{proposition}

Here,  $\stepE{e}{e'}$  is the standard expression evaluation relation, and progress can be proven by induction in the standard way. On the other hand, proving
preservation is more interesting, mainly due to the rules from Fig.~\ref{fig:new_map_typing_rules} for complex data structures. To prove preservation, we first need to prove a lemma that that relates each sum function generated in the refinement of a mapping  to that of the corresponding sum function generated for each entry (value) in the mapping. With such a lemma, we can prove the following preservation theorem in a fairly standard way: 
\begin{proposition}[Preservation for Expressions]
  \label{prop:pres-expr}
  If \ $\tjudg{\PrfCtx}{\expr}{\type}$ and $\stepE{\expr}{\expr'}$, then $\tjudg{\PrfCtx}{\expr'}{\type}$.
\end{proposition}

Next, to formulate the progress and preservation theorems for statements, we first introduce the following notation:
\[
  \begin{array}{lr}
  \store & \text{(State variable store)} \\
  \stepS{s}{\store}{s'}{\store'}{\stmtSub} & \text{(Statement evaluation relation)} \\
  \tjudgStore{\store} & \text{(Store typing)}
  \end{array}
\]
The state variable store $\store$ contains the value bindings for the state variables.
The statement evaluation relation $\stepS{s}{\store}{s'}{\store'}{\stmtSub}$ indicates that, starting with state variable bindings $\store$, the statement $s$ will step to a statement $s'$ with updated bindings $\store'$ and a set of local variable bindings $\stmtSub$ to apply to the statements following $s'$.
The store typing judgment $\tjudgStore{\store}$ asserts that the bindings in $\store$ are well-typed.

With this notation in place, we formulate the progress and preservation theorems for statements as Propositions~\ref{prop:prog-stmt} and~\ref{prop:pres-stmt}.
Note that $\sassume \efalse$ in Proposition~\ref{prop:prog-stmt} corresponds to a failed runtime check.
\begin{proposition}[Progress for Statements]\label{prop:prog-stmt}
  If \ $\tjudgS{\PrfCtxPg}{s}{\Env;\Lock',\Guards}$ and \ $\tjudgStore{\store}$, then $s = \sskip$, or $s = \sassume \efalse$, or there exist $s', \store', \store_{sub}$ such that $\stepS{s}{\store}{s'}{\store'}{\store_{sub}}$.
\end{proposition}

\begin{proposition}[Preservation for Statements]
  \label{prop:pres-stmt}
  If all of the following hold:
  \begin{enumerate}[label=(\alph*)]
  \item \label{eq:pres-stmt-hstmt} 
  $\tjudgS{\PrfCtxS}{s}{\Env;\Lock',\Guards}$
  \item \label{eq:pres-stmt-hstore}
  $\tjudgStore{\store}$
  \item \label{eq:pres-stmt-hstep}
  $\stepS{s}{\store}{s'}{\store'}{\store_{sub}}$
  \end{enumerate}
  then there exist $\PrfGdS', \Env', \Guards'$ such that
  \begin{enumerate}[label=(\roman*)]
  \item \label{eq:pres-stmt-cstmt} 
  $\tjudgS{\epsilon;\PrfGdS'}{s'}{\Env';\Lock',\Guards'}$
  \item \label{eq:pres-stmt-cstore} 
  $\tjudgStore{\store'}$
  \end{enumerate}
\end{proposition}

While progress for statements can be proven in the standard way, preservation is trickier to prove since the resulting statement $s'$ may be typed under different environment and guard predicates than $s$.
To this end, we instead prove a strengthened version of preservation that implies Proposition~\ref{prop:pres-stmt}; we refer the reader to the extended version of the paper \cite{extended-version} for the details.

\section{Type Inference}~\label{sec:infer}

In this section, we describe our algorithm for automatically inferring refinement type annotations. Since writing refinement type annotations for every variable can be  cumbersome,  \toolname infers types for both local and global (state) variables.

As mentioned earlier, the key idea underlying our type inference algorithm is to reduce the type inference problem to Constrained Horn Clause (CHC) solving~\cite{chc}. However, since not all arithmetic operations in the program are guaranteed to be overflow-safe, the CHC constraints generated by the type checking algorithm will, in general, not be satisfiable. Thus, our high-level idea is to treat the overflow safety constraints as  \emph{soft constraints} and find type annotations that try to satisfy as many soft clauses as possible. 

Before we explain our algorithm, we first clarify some assumptions that we make about the type checking phase. First, we assume that, during constraint generation, the refinement type of every variable $v$ without a type annotation is represented with a fresh uninterpreted predicate symbol $p_v$. Second, we assume that the generated constraints are marked as either being hard or soft. In particular,  the overflow safety constraints generated using the {\sc TE-Plus}, {\sc TE-Minus} etc. rules are considered to be soft constraints, while all  other constraints are marked as hard. Third, we assume that the constraint generation phase imposes a partial order on the soft constraints. In particular, if $c, c'$ are soft constraints generated when analyzing expressions $e, e'$ and $e$ must be evaluated before $e'$, then we have $c \prec c'$. As we will see shortly, this partial order allows us to assume, when trying to satisfy some soft constraint $c$, that  overflow safety checks that happened before $c$ did not fail.

Our type inference procedure {\sc InferTypes} is shown in Figure~\ref{alg:infer} and takes three inputs, namely a set of soft and hard constraints, $C_s$ and $C_h$ respectively, and a partial order $\preceq$ on soft constraints as described above. The output of the algorithm is a mapping $\Sigma$ from each uninterpreted predicate symbol to a refinement type annotation.

Initially, the algorithm starts by initializing every uninterpreted predicate symbol $p_v$ to true (line~\ref{line:infer-init-preds}) and then enters a loop where the interpretation for each $p_v$ is gradually strengthened. In particular, in every loop iteration, we pick one of the soft constraints $c$ and try to satisfy $c$ along with all the hard constraints $C_H$. In doing so, we can assume all the overflow safety checks that happened to prior to $c$; thus, at line~\ref{line:infer-solvechc}, we feed the following constraint to an off-the-shelf CHC solver:
\[
\Big ( \Big (\bigwedge_{c_i \prec c} c_i \Big ) \rightarrow c \Big ) \land \bigwedge_{c_j \in C_h} c_j
\]
In other words, we try to satisfy this particular soft constraint together with all the hard constraints, under the assumption that previous overflow checks did not fail. This assumption is safe since \toolname inserts run-time overflow checks for any operation that cannot be statically verified. If this constraint is unsatisfiable, we move on and try to satisfy the next soft constraint. On the other hand, if it is satisfiable, we strengthen the type annotation for the unknown predicates by conjoining it with the assignment produced by the CHC solver (line \ref{line:infer-update-pred}).

\begin{figure*}[!t]
        \begin{algorithmic}[1]
        \Procedure{InferTypes}{$C_s$, $C_h$, $\preceq$}
            \State \textbf{Input:} Soft clauses $C_s$, hard clauses $C_h$, partial order $\preceq$ on $C_s$
            \State \textbf{Output:} Mapping $\Sigma$ from uninterpreted predicates $C_s \cup C_h$ to refinement type annotations
            \vspace{0.05in}
            \State $\mathcal{P} \gets \mathsf{Preds}(C_s \cup C_h)$
            \State $\Sigma \gets \{ p \mapsto \top \ | \ p \in \mathcal{P}\}$ \label{line:infer-init-preds}
            \For{$c \in C_s$}
               \State $\Phi_A \gets \bigwedge_{c_i \prec c} c_i$; $\Phi_H \gets \bigwedge_{c_j \in C_h} c_j$
               \State $(r, \sigma) \gets \mathsf{SolveCHC}((\Phi_A \rightarrow c) \land \Phi_H)$ \label{line:infer-solvechc}
               \If{$r = \mathsf{UNSAT}
               $}
                  \State \textbf{continue}
               \EndIf
               \For{$p \in \mathsf{Dom}(\sigma)$}
                  \State $\phi \gets \Sigma(p) \land \sigma(p)$ \label{line:infer-update-pred}
                  \State $\Sigma \gets \Sigma[p \mapsto \phi]$
               \EndFor
            \EndFor
            \vspace{0.05in}
            \State \Return $\Sigma$
    \EndProcedure
    \end{algorithmic}
  
  
  

  \caption{Type inference procedure}
  \label{alg:infer}
\end{figure*}

\section{Implementation}~\label{sec:impl}

We implemented our type checking and type inference algorithms in a prototype called \toolname.
Given a Solidity source file, \toolname will run the aforementioned algorithms and, for each arithmetic operation, output whether its corresponding soft constraint is satisfied or violated.
Our tool is written in Haskell, with about 3000 lines of code for the frontend (e.g., Solidity preprocessing, lowering, SSA transformation) and 4000 lines of code for the backend (e.g., constraint generation, SMT embedding, solving).
\toolname uses the Z3 SMT solver~\cite{z3} and the Spacer CHC solver \cite{spacer2014} in its backend, and it leverages the Slither~\cite{slither} analyzer as part of its front-end.
In what follows, we  discuss various aspects of  Solidity  and how we handle them in the MiniSol IR.

\paragraph{Preprocessing Solidity code.}
To facilitate verification, we preprocess Solidity smart contracts before we translate them to the MiniSol IR.
In particular, we partially evaluate constant arithmetic expressions such as \lstinline{6 * 10**26}, which frequently appear in  constructors and initialization expressions.
We also inline internal function calls.

\paragraph{Function call handling} Our implementation automatically inserts fetch and commit statements between function call boundaries, such as at the beginning of a function, before and after function calls or calls to external contracts, and before returning from a function.
This includes most function calls, including monetary transfers like \lstinline{msg.sender.send(..)} and calls to "non-pure" functions in the same file. In Solidity, functions may be marked "pure", 
meaning they do not read or write to state variables.
We modified the \textsc{TS-Call} rule to not require the state variables to be unlocked when making calls to a pure function.

\paragraph{Loops} Our implementation supports while-loops and for-loops using a variation of the \textsc{TS-If} rule.
Performing effective type inference in the presence of loops (particularly, doubly-nested loops) is challenging, as existing CHC solvers have a hard time solving the resulting constraints. Inferring invariants for complex loops could be an interesting direction for future work, however, in practice, such loops are rare \cite{mariano2020demystifying} and we found they are mostly irrelevant to proving overflow safety. Our prototype implementation does not support complex control flow inside loops such as break or continue statements.

\paragraph{References and aliasing.}
Since everything is passed by value in MiniSol, the language shown in Figure~\ref{fig:syntax} does not have aliasing.  Thus, as standard ~\cite{ESC-Java,boogie}, our translation from Solidity to MiniSol introduces additional mappings to account for possible aliasing.  At a high-level, the idea is that, for any variables $X = \{x_1, \ldots, x_n\}$ that may alias each other, we introduce a mapping $M_X$ and model stores (resp. loads) to any $x_i$ as writing to (resp. reading from) $M_X[x_i]$.

\paragraph{Unsupported features.}
Our implementation does not support some Solidity features such as inline assembly, states of external contracts, bitwise operations, and exponentiation. When translating to the MiniSol IR, we model these statements using "havoc" expressions as is standard~\cite{boogie}.

\section{Evaluation}~\label{sec:eval}

In this section, we describe a series of experiments that are designed to answer the following research questions:
\begin{itemize}[leftmargin=*]
    \item {\bf RQ1:} Can \toolname successfully remove redundant overflow checks without manual annotations?
  \item {\bf RQ2:} What is \toolname's false positive rate?
    \item {\bf RQ3:} How long does \toolname take to type check real-world smart contracts?
   \item {\bf RQ4:} How does \toolname compare against existing state-of-the art tools?
   \item {\bf RQ5:} How important are type inference and the proposed refinement typing rules for more complex data structures?
\end{itemize}

\paragraph{Baseline.}
To answer our fourth research question, we compare \toolname against \verismart, which is a state-of-the-art Solidity verifier that focuses on proving arithmetic safety properties~\cite{verismart2020}. We choose   \verismart as our baseline because it has been shown to outperform other smart contract verifiers for checking arithmetic safety~\cite{verismart2020}. \verismart infers so-called \emph{transaction invariants} through a domain-specific instantiation of the \emph{counterexample-guided inductive synthesis (CEGIS)} framework~\cite{cegis} and uses the inferred invariants to discharge potential overflow errors. To perform this evaluation, we built the latest version of \verismart from source (version from May 31, 2020) and ran both tools on all the benchmarks, using versions 0.4.26 and 0.5.17 of the Solidity compiler.

\paragraph{Setup}
We performed all experiments on a computer with an AMD Ryzen 5900X CPU and 32GB of RAM.
We used version 4.8.9 of the Z3 SMT solver, which also has the built-in Spacer CHC solver \cite{spacer2014}.  In our experiments, we set a time limit of \zthreeTimeout\ seconds per Z3 query.

\paragraph{Settings} Our implementation can be used in two modes that we refer to as \toolauto and \toolsemi. In particular, \toolauto is fully automated and uses our type inference procedure to infer type annotations for state variables (i.e., contract invariants). On the other hand, \toolsemi requires the programmer to provide contract invariants but automatically infers types for local variables. To use \toolname in this semi-automated mode, we manually wrote refinement type annotations for state variables by inspecting the source code of the contract.

\subsection{Benchmarks}

We evaluate \toolname on two different sets of benchmark suites that we refer to as ``\verismartdata benchmarks'' and ``\etherscandata benchmarks''.
As its name indicates, the former one is taken from the \verismart evaluation \cite{verismart2020} and consists of 60 smart contracts that have at least one vulnerability reported in the CVE database.
Since these benchmarks are from 2018 and do not reflect rapid changes in smart contract development, we also evaluate our approach on another benchmark suite collected from \etherscan.
The latter benchmark suite consists of another 60 contracts randomly sampled from \etherscan in March 2021.
Table~\ref{tbl:ether-stat} presents some relevant statistics about the contracts in this dataset.  
Since the goal is to discharge redundant \safemath\footnote{Recall that \safemath is a library that inserts runtime checks before each arithmetic operation.} calls, we preprocessed the benchmarks by replacing all \safemath method calls with their equivalent unchecked operations.
Thus, the reader should be advised that the "\verismartdata benchmarks" are not exactly the same as the ones in \cite{verismart2020}.

Although our preprocessed benchmarks technically do not contain any \safemath runtime checks, in the following discussion, we will use the terms "run-time checks", "redundant checks", and "necessary checks" to refer to arithmetic operations, provably overflow-safe operations, and operations whose overflow-safety cannot be proven, respectively.

\begin{table}
\caption{Statistics about our benchmarks from \etherscan}
\label{tbl:ether-stat}
\begin{tabular}{|l|c|}
\hline
{\bf Description } & {\bf Stats}     \\ \hline \hline
\hline
\% contracts containing  mappings of structs & 26.7\% \\ \hline
\% contracts containing  nested mappings  & 81.3\%  \\ \hline
Average lines of code & \lineCntAvg  \\ \hline
Average number of methods & 29.6  \\
\hline
\end{tabular}
\end{table}

\subsection{Discharging Redundant SafeMath Calls}

We performed a manual inspection of all 120 benchmarks to determine how many of the \safemath run-time checks are redundant. In total, among the \TotTotOps\ run-time checks, we determined  \TotTotGtNeg\ of them to be redundant (\VsEvalGtNeg\ in \verismartdata and \EsRedundant\ in \etherscandata).  In this experiment, we evaluate what percentage of these redundant checks \verismart, \toolauto, and \toolsemi are able to discharge.

The results from this evaluation are shown in  Table~\ref{tbl:check}. The key-take away is that \toolauto is able to discharge more run-time checks in both benchmark suites. In particular, for the \verismartdata benchmarks, \toolauto can discharge approximately \VsEvalSafePctDelta\% more overflow checks, and for the \etherscandata benchmarks, this difference increases to \EsEvalSafePctDelta\%. If we additionally leverage manually-written refinement type annotations, then \toolsemi can discharge almost \TotSsSafePctRounded\% of the redundant checks across both datasets. \\

\begin{table}
\caption{{Number and percentage of redundant  overflow checks that can be eliminated by each tool. The Ops column displays the total number of SafeMath checks. Under each tool, the Safe column lists the number of arithmetic operations that can be proven safe by that tool, and the \% column shows the percentage of redundant overflow checks that can be discharged. }}
\label{tbl:check}
\begin{tabular}{|l|l|l|l|l|l|l|l|l|}
\hline
\multirow{2}{*}{Dataset} & \multirow{2}{*}{Ops} & \multirow{2}{*}{\#redundant} & \multicolumn{2}{l|}{\verismart} & \multicolumn{2}{l|}{\toolauto} & \multicolumn{2}{l|}{\toolsemi} \\ \cline{4-9 } 
                      & & & Safe           & \%            & Safe         & \%          & Safe          & \%           \\ \hline
\verismartdata & \VsEvalOps & \VsEvalGtNeg                     & \VsEvalVsSafe            & \VsEvalVsSafePct \%          & \VsEvalAsSafe          & \VsEvalAsSafePct\%        & \VsEvalSsSafe           & \VsEvalSsSafePct\%         \\ \hline
\etherscandata & \totOps & \EsRedundant                    & \vsSafe            & \vsSafePct \%          & \asSafe          & \asSafePct\%        & \ssSafe           & \ssSafePct\%         \\ \hline\hline
\textbf{Total} & \TotTotOps & \TotTotGtNeg & \TotVsSafe & \TotVsSafePct\% & \TotAsSafe & \TotAsSafePct\% & \TotSsSafe & \TotSsSafePct\% \\ \hline
\end{tabular}
\end{table}

\fbox{
\begin{minipage}{0.9\linewidth}
{\bf Result for RQ1:}
\toolname can automatically prove that \TotAsSafePct\% of the redundant \safemath calls  are unnecessary. In contrast, \verismart can only discharge \TotVsSafePct\%.
\end{minipage}
}

\subsection{False Positive Evaluation}

\begin{table}
\caption{Comparison of false positive rates}
\label{tbl:fp}
\begin{tabular}{|l|c|c|c|}
\hline
   & \verismart & \toolauto & \toolsemi \\ \hline
 {\bf Verismart Benchmarks }          &       &            &         \\
\hline
\# false positives & \VsEvalVsFpCount         & \VsEvalAsFpCount     & \VsEvalSsFpCount      \\ \hline
\# true positives & \VsEvalVsTpCount         & \VsEvalAsTpCount     & \VsEvalAsTpCount     \\ \hline
False positive rate & \VsEvalVsFpr\%         & \VsEvalAsFpr\%     & \VsEvalSsFpr\%      \\ \hline
 {\bf Etherscan Benchmarks }          &       &            &         \\
\hline
\# false positives & \vsFpCount         & \asFpCount     & \ssFpCount      \\ \hline
\# true positives & \vsTpCount         & \asTpCount     & \asTpCount     \\ \hline
False positive rate & \vsFpr\%         & \asFpr\%     & \ssFpr\%      \\ \hline
 {\bf Overall }          &       &            &         \\
\hline
\# false positives & \TotVsFpCount & \TotAsFpCount     & \TotSsFpCount      \\ \hline
\# true positives & \TotVsTpCount         & \TotAsTpCount     & \TotSsTpCount     \\ \hline
False positive rate & \TotVsFpr\% & \TotAsFpr\%     & \TotSsFpr\%      \\ \hline
\end{tabular}
\end{table}

Next, we evaluate \toolname's false positive rate in both the fully-automated and semi-automated modes and compare it against \verismart. In particular, Table~\ref{tbl:fp} shows the number of true and false positives  as well as the false positive rate for each tool for both benchmark suites. Across all benchmarks, \verismart reports \TotVsFpCount\ false alarms, which corresponds to a false positive rate of \TotVsFpr\%. On the other hand, \toolauto only reports \TotAsFpCount\ false alarms with a false positive rate of \TotAsFpr\%. Finally, \toolsemi has an ever lower false positive rate of \TotSsFpr\%.

It is worth noting that the false positive rate for \toolsemi can be further reduced by spending additional effort in strengthening our refinement type annotations. When performing  this evaluation, we did not refine our initial type annotations based on feedback from the type checker.

\paragraph{Cause of false positives for \toolauto}
As expected, the cause of most false positives for \toolauto is due to the limitations of the CHC solver. 
There are several cases where the CHC solver times-out or returns unknown, causing \toolauto to report false positives.

\paragraph{Qualitative comparison against \verismart} We believe that \toolname outperforms \verismart largely due to its ability to express relationships between integers and aggregate properties of data structures. In particular, while \verismart can reason about summations over basic mappings, it cannot easily express aggregate properties of more complex data structures.  \\

\fbox{
\begin{minipage}{0.9\linewidth}
{\bf Result for RQ2 and RQ4.}
In its fully automated mode, \toolname has a false positive rate of \TotAsFpr\%. In contrast, \verismart has a significantly higher false positive rate of \TotVsFpr\%.
\end{minipage}
}

\subsection{Running Time}

Next, we investigate the running time of \toolname and compare it against \verismart. Table~\ref{tbl:perf} gives statistics about the running time of each tool. As expected, \toolsemi is faster than \toolauto, as it does not need to rely on the CHC solver to infer contract invariants. However, even \toolauto is significantly faster than \verismart despite producing fewer false positives. \\

\fbox{
\begin{minipage}{0.9\linewidth}
{\bf Result for RQ3 and RQ4:}
In its fully automated mode, \toolname takes an average of \TotAsAvgTime\ seconds to analyze each benchmark, and is significantly faster compared to \verismart despite generating fewer false positives.
\end{minipage}
}

\begin{table}
\caption{Comparison of average running times in seconds}
\label{tbl:perf}
\begin{tabular}{|l|c|c|c|}
\hline
       Dataset & \verismart & \toolauto & \toolsemi \\ \hline
\hline
{\verismartdata}   & \VsEvalVsAvgTime & \VsEvalAsAvgTime & \VsEvalSsAvgTime \\ \hline
{\etherscandata}   & \vsAvgTime & \asAvgTime & \ssAvgTime \\ \hline
\hline
{\bf Overall}    & \TotVsAvgTime & \TotAsAvgTime & \TotSsAvgTime \\ \hline
\end{tabular}
\end{table}

\subsection{Impact of Type Inference}
In this section, we perform an experiment to assess the impact of the type inference procedure discussed in Section~\ref{sec:infer}. Towards this goal, we consider the following two ablations of \toolname:

\begin{table}
\caption{Experiments to evaluate type inference. This experiment is conducted on the \etherscandata benchmarks.}
\label{tbl:ablation-infer}
\begin{tabular}{|l|c|c|c|}
\hline
       & \toolname-NoInfer & \toolname-NoSoft & \toolname \\ \hline
False positive rate    & \AblNoInferFpr\% & \AblNoSoftFpr\% & \asFpr\% \\ \hline
\end{tabular}
\end{table}

\begin{itemize}[leftmargin=*]
    \item {\bf \toolname-NoInfer:} This is a variant of \toolname that does not perform global type inference to infer contract invariants. In particular, \toolname-NoInfer uses \texttt{true} as the type refinement of all state variables; however, it still performs local type inference.
    
    \item {\bf \toolname-NoSoft:} This variant of \toolname differs from the type inference procedure in Fig.~\ref{alg:infer} in that it treats all overflow checks as hard constraints. 
\end{itemize}

In this experiment, we compare the false positive rate of each ablated version against \toolname on the \etherscandata benchmarks. The results of this experiment are summarized in Table~\ref{tbl:ablation-infer}. 

\paragraph{Importance of contract invariant inference.} As we can see by comparing \toolname against \toolname-NoInfer, global type inference is quite important. In particular, if we do not infer contract invariants, the percentage of false positives increases from \asFpr\% to \AblNoInferFpr\%. 

\paragraph{Importance of soft constraints.} Next, we compare the false positive rate of \toolname against that of \toolname-NoSoft. Since \toolname-NoSoft treats all overflow checks as hard constraints, type inference fails if \emph{any} potential overflow in the contract cannot be discharged. Since most contracts contain at least one unsafe overflow, \toolname-NoSoft fails for most benchmarks, meaning that all overflows are considered as potentially unsafe. Thus, the false positive rate of \toolname-NoSoft jumps from \asFpr\% to~\AblNoSoftFpr\%.  

\subsection{Impact of the Type Checking Rules from Section~\ref{sec:agg-props}}
In this section, we describe an ablation study to assess the impact of the more complex typing rules  from Fig.~\ref{fig:new_map_typing_rules} for non-trivial mappings. Towards this goal, we consider the following ablation:

\begin{table}
\caption{Ablation study to evaluate rules from Section~\ref{sec:agg-props}. This experiment is conducted on \etherscan benchmarks that contain non-trivial mappings.}
\label{tbl:ablation-map}
\begin{tabular}{|l|c|c|}
\hline
       & \toolname-NoNested  & \toolname \\ \hline
False positive rate    & \AblNoNestedFpr\%  & \AblNestedFpr\% \\ \hline
\end{tabular}
\end{table}

\begin{itemize}[leftmargin=*]
    \item {\bf \toolname-NoNested:} This variant of \toolname uses the simpler refinement type checking rules from Section~\ref{sec:check-basic} but it does not utilize the typing  rules from Section~\ref{sec:agg-props} that pertain to complex data structures. 
        
\end{itemize}

In this experiment, we compare \toolname-NoNested against \toolname on  those contracts from \etherscan that contain complex data structures.  As shown in Table~\ref{tbl:ablation-map}, the more  involved typing rules from Section~\ref{sec:agg-props} are quite important for successfully discharging overflows in contracts with non-trivial mappings. In particular, without our refinement templates from Section~\ref{sec:agg-props}, the false positive rate increases from \AblNestedFpr\% to \AblNoNestedFpr\% on these benchmarks. \\

\fbox{
\begin{minipage}{0.9\linewidth}
{\bf Result for RQ5:}
Our  proposed  typing rules  from Section~\ref{sec:agg-props} for more complex data structures and the proposed type inference algorithm from Section~\ref{sec:infer} are both important for achieving good results. 
\end{minipage}
}

\section{Limitations}

In this section, we discuss some of the limitations of both our type system as well as prototype implementation.
First, while logical qualifiers in \soltype express relationships between integers and aggregate properties of mappings, they do not allow quantified formulas. In principle, there may be situations that necessitate quantified refinements to discharge arithmetic safety; however, this is not very common.
Second, \toolname uses an off-the-shelf CHC solver to infer refinement type annotations. Since this problem is, in general, undecidable, the CHC solver may return unknown or fail to terminate in a reasonable time. In practice, we set a small time limit of \zthreeTimeout\ seconds per call to the CHC solver. {Third, \toolname is designed with checking arithmetic overflows in mind, so the proposed refinement typing rules may not be as effective for checking other types of properties.}

\section{Related Work}
Smart contract correctness has received significant attention from the programming languages, formal methods, and security  communities in recent years.  In this section, we discuss prior work on refinement type systems and smart contract security.

\paragraph{Refinement types}
Our type system is largely inspired by and shares similarities with prior work on liquid types~\cite{liquidtype08,liquidtype10,liquidhaskell14,rsc2016}. For instance, our handling of temporary contract violations via fetch/commit statements is similar to a simplified version of the \texttt{fold} and \texttt{unfold} operators used in \cite{liquidtype10}. However, a key novelty of our type system is its ability to express and reason about relationships between integer variables and aggregate properties of  complex data structures (e.g., nested maps over struct) that are quite common in smart contracts.  Also, similar to the liquid type work, our system also performs type inference; however, a key novelty in this regard is the use of soft constraints to handle programs where not all arithmetic operations can be proven safe. 

\paragraph{Smart contract verification}
Many popular security analysis tools for smart contracts are based on symbolic
execution~\cite{symbolic-e}. Well-known tools include Oyente~\cite{oyente},
Mythril~\cite{mythril} and Manticore~\cite{manticore}, and they look for 
an execution path that violates a given property or assertion. 
In contrast to these tools, our proposed approach is based on a refinement type system and offers stronger guarantees compared to bug finding tools for smart contracts.

To bypass the  scalability issues associated with symbolic execution,
researchers have also investigated sound and scalable static
analyzers~\cite{ecf,securify,madmax,zeus}. Both Securify~\cite{securify} and
Madmax~\cite{madmax} are based on abstract interpretation~\cite{CousotC77},
which does not suffer from the path explosion problem. ZEUS~\cite{zeus} translates Solidity code into the LLVM IR and 
uses an off-the-shelf verifier to check the specified policy~\cite{smack}. The ECF~\cite{ecf} system is
designed to specifically detect the DAO vulnerability.  In contrast to these techniques, the focus of our work is on proving arithmetic safety, which is a prominent source of security vulnerabilities.

Similar to \toolname, \verismart~\cite{verismart2020} is also (largely) tailored towards arithmetic properties of smart contracts. In particular, \verismart leverages a CEGIS-style algorithm for inferring contract invariants.  As we show in our experimental evaluation, \toolname outperforms \verismart in terms of false positive rate as well as running time. 


Some systems~\cite{Hirai17,GrishchenkoMS18,kframework,verx2019} for reasoning about smart contracts rely on formal verification. These systems typically prove security properties of smart
contracts using existing interactive theorem provers~\cite{coq}, or leverage temporal verification for checking functional correctness~\cite{verx2019}. They
typically offer strong guarantees that are crucial to smart contracts. However,
unlike our system, all of them require significant manual effort to encode the
security properties and the semantics of smart contracts.
Furthermore, \textsc{Solythesis}~\cite{solythesis} inserts runtime checks into smart contract source code to revert transactions that violate contract invariants. We note that this approach is complementary to static verifiers such as \toolname.

\paragraph{Verifying overflow/underflow safety}
The problem of checking integer overflow/underflow in software systems is a well-studied problem in the verification community. In particular,
Astree~\cite{astree} and Sparrow~\cite{sparrow} are both based on abstract interpretation~\cite{CousotC77} and are tailored towards safety-critical code such as avionics software. As we argue throughout the paper, verifying arithmetic safety of smart contracts requires  reasoning about aggregation properties over mappings; hence, standard numeric abstract domains such as the ones used in Astree would not be sufficient for discharging over- and under-flows in Solidity programs.

\paragraph{CHC Solving}
We are not the first to use CHC solvers in the context of refinement type checking. For example, Liquid Haskell \cite{liquidhaskell14} uses a predicate-abstraction based horn clause solver for refinement type inference. However, in contrast to our work, Liquid Haskell does not tackle the problem of MaxCHC type inference, which is critical for discharging the maximum number of runtime overflow checks in smart contracts.

To the best of our knowledge, the only prior work that can potentially address our MaxCHC problem is in the context of network repair. In particular, 
\cite{hojjat2016optimizing} formulate the problem 
of repairing buggy SDN configurations as an optimization problem over constrained Horn clauses. They generalize their proposed method to Horn clause optimization over different types of lattices, and our type inference algorithm may be viewed as an instantiation of their more general framework. However, in contrast to their approach, our algorithm leverages domain-specific observations for efficient (but potentially suboptimal) type inference in the context of smart contracts. In particular, we use the fact that all soft constraints correspond to overflow checks that are guaranteed to be satisfied (either because they are safe or we insert a runtime check) to check each soft constraint independently.

\section{Conclusion}~\label{sec:concl}
 We have presented \soltype, a refinement type system for Solidity that can be used to prove the safety of arithmetic operations. Since integers in smart contracts often correspond to financial assets (e.g., tokens), ensuring the safety of arithmetic operation is particularly important in this context. One of the distinguishing features of our type system is its ability to express and reason about  arithmetic relationships between integer variables and aggregations over complex data structures such as multi-layer mappings. We have implemented our proposed type system in a prototype called \toolname, which also has type inference capabilities. 
 We have evaluated \toolname on {\TotContractCount}\ smart contracts from 
 two datasets and demonstrated that it can fully automatically discharge \TotAsSafePct\% of redundant \safemath calls in our benchmarks. Furthermore, \toolname, even when used in a fully automated mode, significantly outperforms \verismart, a state-of-the-art Solidity verifier, in terms of both false positive rate and running time.
 
While the design of \soltype was largely guided from the perspective of proving arithmetic safety, our proposed type system could also be used for proving other types of properties. In future work, we plan to explore the applicability of our refinement type system in other settings and extend it where necessary.
\begin{acks}
We thank the anonymous reviewers, as well as our shepherd James R. Wilcox, for their helpful comments and valuable suggestions.

This work was partially supported by NSF Grants 2027977, 1908494, 1811865, 1762299, and the Google Faculty Research Award.
\end{acks}

\bibliography{main}

\ifextended
\appendix

\input{appendix/soundness-macros.sty}

In this appendix, we present the proof of the progress and preservation theorems.
First, we define the expression evaluation and typing rules in Section~\ref{sec:appendix-expr-rules}.
We then prove several useful expression lemmas, progress of expressions, and preservation of expressions in Section~\ref{sec:appendix-expr-proofs}.
Sections~\ref{sec:appendix-stmt-rules} and~\ref{sec:appendix-stmt-proofs} are analogous.

For the purposes of the proofs, we consider the subset of the expression language without 1) multiplication or division; and 2) without logical operators such as $\land, \lor$ (logical operators are still included as refinements terms, however).

\section{Expression Rules} 
\label{sec:appendix-expr-rules}

We first define values, expression evaluation contexts, and the expression evaluation relation, which have the standard meanings.
\[
\begin{array}{r c l l}
\val & ::= & &\textbf{Value} \\
&& n \mid \etrue \mid \efalse \mid \\
&\mid& \emapping{\btype}{\overrightarrow{n_i \mapsto \val_i}} \mid \estruct{S}{\overrightarrow{x_i \mapsto \val_i}} \\
\Ectx & ::= & & \textbf{Expression evaluation context} \\
&& \cdot \mid \Ectx \binop \expr_2 \mid \val \binop \Ectx &  \\ 
& \mid & \eind{\Ectx}{e_2} \mid \eind{\val_1}{\Ectx} \mid \eupd{\Ectx}{\expr_2}{\expr_3}
         \mid \eupd{\val_1}{\Ectx}{\expr_3} 
         \mid \eupd{\val_1}{\val_2}{\Ectx} 
       & \text{mapping index/update} \\
& \mid & \eind{\Ectx}{.x} \mid \eupd{\Ectx}{.x}{\expr} \mid \eupd{\val}{.x}{\Ectx} & \text{struct index/update} \\
\\
\multicolumn{3}{c}{\stepE{\expr}{\expr'}} & \textbf{Expression evaluation relation}
\end{array}
\]

\subsection{Expression Rules}

We now present the operational semantics for expressions. First, we define $\mathsf{ZeroVal}(\btype)$, the \emph{zero value} of a base type $\btype$, similar to how it is defined in Solidity:

\[
\ZeroVal{\btype} = \begin{cases}
  0 & \text{ if } \btype = \tint \\
  \efalse & \text{ if } \btype = \tbool \\
  \emapping{\btype}{} & \text{ if } \btype = \tmapping{\btype} \\
  \estruct{S}{\overrightarrow{x_i \mapsto \ZeroVal{\btype_i}}} & \text{ if } \btype = \tstruct{S} \text{ and } \forall x_i. \defns(S, x_i) = \btype_i
\end{cases}
\]

The evaluation rules are shown in Figure~\ref{fig:expr-eval-rules}. Here we subscript binary operations with $\mathbb{N}$ to indicate an operation performed in the $\mathbb{N}$ domain instead of as a syntactic object.

\begin{figure}
\begin{mathpar}
\inferrule*[right=EE-Ctx]{
  \stepE{\expr}{\expr'}
}{
  \stepE{\Ectx(\expr)}{\Ectx(\expr')}
}
\and
\inferrule*[right=EE-Plus]{
  n = n_1 +_\mathbb{N} n_2
}{
  \stepE{n_1 + n_2}{n}
}
\and
\inferrule*[right=EE-Minus]{
  n_1 \geq_\mathbb{N} n_2 \\
  n = n_1 -_\mathbb{N} n_2
}{
  \stepE{n_1 - n_2}{n}
}
\\
\inferrule*[right=EE-Rel]{
  b = n_1 \binop_\mathbb{N} n_2 \\
  \binop \in \{=, \neq, \geq, \leq, >, <\}
}{
  \stepE{n_1 \binop n_2}{b}
}
\\
\inferrule*[right=EE-MapInd1]{
  v_M = \emapping{\btype}{n_1 \mapsto \val_1, \dots, n_i \mapsto \val_i, \dots, n_k \mapsto \val_k}
}{
  \stepE{
    \eind{v_m}{n_i}
  }{\val_i}
}
\\
\inferrule*[right=EE-MapInd2]{
  \not\exists i.\ n_i = n \\
}{
  \stepE{
    \eind{\left(\emapping{\btype}{n_1 \mapsto \val_1, \dots, n_k \mapsto \val_k}\right)}{n}
  }{\ZeroVal{T}}
}
\\
\inferrule*[right=EE-MapUpd1]{
  v_M = \emapping{\btype}{n_1 \mapsto \val_1, \dots, n_i \mapsto \val_i, \dots, n_k \mapsto \val_k}
}{
  \stepE{
    \eupd{v_M}{n_i}{\val}
  }{
    \emapping{\btype}{n_1 \mapsto \val_1, \dots, n_i \mapsto \val, \dots, n_k \mapsto \val_k}
  }
}
\\
\inferrule*[right=EE-MapUpd2]{
  v_M = \emapping{\btype}{n_1 \mapsto \val_1, \dots, n_k \mapsto \val_k} \\
  \not\exists i.\ n_i = n
}{
  \stepE{
    \eupd{v_M}{n}{\val}
  }{
    \emapping{\btype}{n_1 \mapsto \val_1, \dots, n_k \mapsto \val_k, n \mapsto \val}
  }
}
\\
\inferrule*[right=EE-SctInd]{ }{
  \stepE{
    \eind{\left(\estruct{S}{x_1 \mapsto \val_1, \dots, x_i \mapsto \val_i, \dots, x_k \mapsto \val_k}\right)}{.x_i}
  }{\val_i}
}
\\
\inferrule*[right=EE-SctUpd]{
  v_R = \estruct{S}{x_1 \mapsto \val_1, \dots, x_i \mapsto \val_i, \dots, x_k \mapsto \val_k}
}{
  \stepE{
    \eupd{v_R}{.x_i}{\val}
  }{\estruct{S}{x_1 \mapsto \val_1, \dots, x_i \mapsto \val, \dots, x_k \mapsto \val_k}}
}
\end{mathpar}
\caption{The expression evaluation rules.}
\label{fig:expr-eval-rules}
\end{figure}

All of these rules are standard with the exception of the mapping rules. The \textsc{EE-MapInd1} rule retrieves the accessed value if it is contained in the mapping; otherwise, the \textsc{EE-MapInd2} rule returns the zero value of the value type. Similarly, \textsc{EE-MapUpd1} replaces the currently existing entry if the key exists; otherwise, the key is added to the mapping in \textsc{EE-MapUpd2}.

\subsection{Template Pairings}

We introduce new, less ambiguous notation for constructing the templates used for the \textsc{TE-MapInd} and \textsc{TE-MapUpd} rules.
A \emph{template pairing} is a function $\Hfam$ that takes a pair $\btype, \btype_h$ and produces a set of tuples of templates of sort $\btype$  that can be synthesized using holes of type $\tmapping{\btype_h}, \btype_h$ (resp.), along with the access path $w$ used.
\[
\Hfam(\btype, \btype_h) = \{(\hole_1, \hole_2, w) \mid 
  \tjudgP{\tmapping{\btype_h}}{\hole_1}{\btype}{w} \text{ and }
  \tjudgP{\btype_h}{\hole_2}{\btype}{w}
\}
\]
Observe that for any given pair of $\btype$ and $\btype_h$, $\Hfam(\btype, \btype_h)$ will be finite, so by convention we assume that there exists a unique ordering
\[
\mathcal{H}(T, T_h) = \{(\hole_{11}, \hole_{12}, w_1), \dots, (\hole_{n1}, \hole_{n2}, w_n)\}
\]

\subsection{New Notation in TE-MapInd and TE-MapUpd Rules}

Using the template pairing notation, we update the \textsc{TE-MapInd} and \textsc{TE-MapUpd} rules:
\begin{mathpar}
\inferrule*[Right=TE-MapInd]{
  \tjudg{\Env;\Guards}{e_1}{\tmapping{\btype}}
  \and
  \tjudg{\Env;\Guards}{e_2}{\tint} \\
  \Hfam(\tint, \btype) = \{(\hole_{i1}, \hole_{i2}, w_i), \dots, (\hole_{\ell 1}, \hole_{\ell 2}, w_\ell)\} \\
  \qual_a = \bigwedge_{i=1}^\ell \hole_{i2}(\nu) \leq \hole_{i1}(e_1)
}{
  \tjudg{\Env;\Guards}{e_1[e_2]}{\rtype{\btype}{\nu = e_1[e_2] \land \qual_a}}
}
\\
\inferrule*[Right=TE-MapUpd]{
  \tjudg{\Env;\Guards}{e_1}{\tmapping{\btype}} \and
  \tjudg{\Env;\Guards}{e_2}{\tint} \and
  \tjudg{\Env;\Guards}{e_3}{\btype} \\
  \Hfam(\tint, \btype) = \{(\hole_{i1}, \hole_{i2}, w_i), \dots, (\hole_{\ell 1}, \hole_{\ell 2}, w_\ell)\} \\
  \qual_a = \bigwedge_{i=1}^\ell \left(\hole_{i1}(\nu) = \hole_{i1}(e_1) - \hole_{i2}(e_1[e_2]) + \hole_{i2}(e_3)\right)
}{
  \tjudg{\Env;\Guards}{\eupd{e_1}{e_2}{e_3}}{\rtype{\tmapping{\btype}}{\nu = \eupd{e_1}{e_2}{e_3} \land \qual_a}}
}
\end{mathpar}

\subsection{Typing Rules Omitted in the Main Text}

We omitted some typing rules in the main text, notably that of the constants, so we provide them in Figure~\ref{fig:expr-typing-rules-omitted}.

\begin{figure}
\begin{mathpar}
\inferrule*[right=TE-Nat]{
  0 \leq_\mathbb{N} n \leq_\mathbb{N} \pmaxint
}{
  \tjudg{\Env; \Guards}{n}{\rtype{\tint}{\nu = n}}
}
\and
\inferrule*[right=TE-True]{ }{
  \tjudg{\Env; \Guards}{\etrue}{\rtype{\tbool}{\nu = \etrue}}
}
\\
\inferrule*[right=TE-False]{ }{
  \tjudg{\Env; \Guards}{\efalse}{\rtype{\tbool}{\nu = \efalse}}
}
\and
\inferrule*[right=TE-Rel]{
  \tjudg{\Env; \Guards}{e_1}{\tint} \\
  \tjudg{\Env; \Guards}{e_2}{\tint} \\
  \\\\
  \binop \in \{=, \neq, \geq, \leq, >, <\}
}{
  \tjudg{\Env; \Guards}{e_1 \binop e_2}{\rtype{\tbool}{\nu = e_1 \binop e_2}}
}
\\
\inferrule*[right=TE-StructConst]{
  \defns(S) = \{x_1 : \btype_1, \dots, x_n : \btype_n\}, \and
  \tjudg{\Env; \Guards}{\val_i}{\btype_i} \\
  \val = \estruct{S}{x_1 \mapsto \val_1, \dots, x_n \mapsto \val_n}
}{
  \tjudg{\Env; \Guards}{\val}{\rtype{\tstruct{S}}{\nu = v}}
}
\\
\inferrule*[right=TE-MappingConst]{
  \text{for $i = 1 \dots k$,} \and \tjudg{\Env; \Guards}{\val_i}{\btype} \\\\
  \Hfam(\tint, \btype) = \{(\hole_{11}, \hole_{12}, w_1), \dots, (\hole_{\ell 1}, \hole_{\ell 2}, w_\ell)\} \\\\
  \val = \emapping{\btype}{n_1 \mapsto \val_1, \dots, n_k \mapsto \val_k}
  \\
  \phi = \bigwedge_{j=1}^\ell \hole_{j1}(\nu) = \ConstAgg(\btype, \hole_{j1}, w_j, \val)
}{
  \tjudg{\Env; \Guards}{\val}{
    \rtype{\tmapping{\btype}}{\nu = v \land \phi}
  }
}
\end{mathpar}
    \caption{The expression typing rules omitted in the main text.}
    \label{fig:expr-typing-rules-omitted}
\end{figure}

The \textsc{TE-True}, \textsc{TE-False}, and {\sc TE-Rel} rules are standard. The \textsc{TE-StructConst} rule directly refines the type of a struct constant with its exact concrete value, since a struct can be directly encoded into SMT in our chosen theory.
The \textsc{TE-Nat} rule is similar to what would be an integer typing rule in a typical type system, except that it explicitly enforces the number to be within machine bounds. Note that this only affects numbers appear in the program and does not necessarily include numbers appearing in refinements.

The most interesting rule is \textsc{TE-MappingConst}. Similar to \textsc{TE-StructConst}, the type of a mapping constant is refined with its exact concrete value (since the number of values is finite), but further includes the clause $\phi$ that sets the values of the synthesized symbolic sum functions. Specifically, $\phi$ equates the symbolic value $\hole_{j1}(\nu)$ representing the aggregation to the concrete, numerical value of the aggregation computed by the $\ConstAgg$ helper function.

The helper functions are defined as follows:
\begin{align*}
    \ConstAgg(\btype, \hole, w_j, \val) &= \begin{cases}
    \bot & \text{if } \hole = \holet \\
    \ConstSum(w, \val) & \text{if } \hole = \psum(\hole') \text{ and } \btype = \tint
    \end{cases} \\
    \ConstSum(w, \val) &= \begin{cases}
    n & \text{if } \val = n \\
    \sum_{i=1}^k \ConstSum(w, v_i) & \text{if } \val = \emapping{\btype}{\overrightarrow{n_i \mapsto \val_i}} \text{ for } i = 1 \dots k \\
    \ConstSum(w', \val') & \text{if } w = xw' \text{ and } v = \estruct{S}{\dots, x \mapsto v', \dots}
    \end{cases}
\end{align*}

$\ConstAgg(\btype, \hole, w, v)$ computes the concrete value of the aggregation of $v$, where the aggregated values are only those obtained by accessing the struct fields in $w$. In our presentation, we only have the sum aggregation, which is computed by the $\ConstSum$ helper function. $\ConstSum(w, v)$ computes the concrete sum of a number, mapping, or struct field in the natural number domain.

\begin{example}
Suppose $v_i = \estruct{S}{x_a \mapsto i, x_b \mapsto 1}$ for $i = 1, \dots, 10$. Then if $v = \emapping{\tstruct{S}}{\overrightarrow{i \mapsto v_i}}$, then
\begin{align*}
    \ConstSum(x_a, v) &= 55 \\
    \ConstSum(x_b, v) &= 10
\end{align*}
\end{example}

\section{Progress and Preservation for Expressions}
\label{sec:appendix-expr-proofs}

\subsection{Some useful lemmas}

\begin{lemma}[Zero values are well-typed]
    For every $\btype$, there exists a $\type$ such that $\tjudg{\Env;\Guards}{\mathsf{ZeroVal}(\btype)}{\type}$ and $\subty{\Env;\Guards}{\type}{\btype}$.
\end{lemma}
\begin{proof} By induction on $\btype$.
\end{proof}

\begin{lemma}[Canonical forms]
    \label{lem:canonical-forms}
    If $\tjudg{\Env;\Guards}{\val}{\rtype{\btype}{\qual}}$, then
    \begin{enumerate}[label=(\roman*)]
    \item If $\btype = \tint$, then $v \in \mathbb{N}$.
    \item If $\btype = \tbool$, then $v = \etrue$ or $v = \efalse$.
    \item If $\btype = \tmapping{\btype'}$, then $v = \emapping{\btype}{\overrightarrow{n_i \mapsto v_i}}$.
    \item If $\btype = \tstruct{S}$, then $v = \estruct{S}{\overrightarrow{x_i \mapsto v_i}}$ where $\defns(S) = \{x_1 : \btype_1, \dots, x_n : \btype_n \}$ and $i = 1 \dots n$.
    \end{enumerate}
\end{lemma}
\begin{proof} By induction on the derivation of $\tjudg{\Env;\Guards}{\val}{\rtype{\btype}{\qual}}$.
\end{proof}

\begin{lemma}[Inversion for evaluation context type]
  \label{lem:eval-ctx-arg-typed}
  If $\tjudg{\Env;\Guards}{\Ectx(\expr)}{\type}$, then there exists a $\type'$ such that $\tjudg{\Env;\Guards}{\expr}{\type'}$.
\end{lemma}
\begin{proof} By induction on $\Ectx$ and inverting the typing judgment to apply the inductive hypothesis.
\end{proof}

\begin{lemma}[Well-typedness of swapping evaluation context argument]
  \label{lem:eval-ctx-swap-typed}
  If $\tjudg{\Env;\Guards}{\Ectx(\expr)}{\type}$ and $\tjudg{\Env;\Guards}{\expr}{\type'}$ and $\tjudg{\Env;\Guards}{\expr'}{\type'}$, then $\tjudg{\Env;\Guards}{\Ectx(\expr')}{\type}$.
\end{lemma}
\begin{proof} By induction on $\Ectx$. In particular, since $\Ectx(e)$ is well typed, either $\Ectx = \cdot$ (in which case the result follows immediately) or there exists some well-typed $\Ectx'(e)$ that is a subexpression of $\Ectx(e)$, in which case the required result can be derived via an application of the inductive hypothesis.
\end{proof}

\begin{lemma}[Canonical form of a boolean value]
  \label{lem:canonical-bool-value}
  If $b \in \{\etrue, \efalse\}$ and $\tjudg{\epsilon;\epsilon}{\val}{\rtype{\tbool}{\qual}}$ and $\Encode(\qual) \implies \Encode(\nu = b)$, then $\val = b$.
\end{lemma}
\begin{proof} By induction on $\tjudg{\epsilon;\epsilon}{\val}{\rtype{\tbool}{\nu = b}}$.
The only relevant cases are {\sc TE-True} (trivial), {\sc TE-False} (trivial), or {\sc TE-Sub}.
The latter case follows immediately by transitivity of $\implies$ and an application of the inductive hypothesis.
\end{proof}

\begin{lemma}[Non-identity struct holes must use field access first]
\label{lem:struct-holes}
If $\tjudgP{\tstruct{S}}{\hole}{\btype}{w}$ and $\hole \neq \holet$, then there exist $\hole', w', x, \btype_h$ such that
\begin{align}
    \tjudgP{\btype_h&}{\hole'}{\btype}{w'} \\
    w &= xw' \\
    \defns(S, x) &= \btype_h \\
    \hole'(\holet[.x]) &= \hole
\end{align}
\end{lemma}
\begin{proof} By induction on the derivation of $\tjudgP{\tstruct{S}}{\hole}{\btype}{w}$.
\begin{itemize}
\item Case {\sc TH-Hole}: Impossible.

\item Case {\sc TH-Fld}: Then
\begin{align}
    H &= \pfld_{x_1}(\hole_1) \\
    \btype &= \tmapping{\btype'} \\
    w &= w_1 x_1 \\
    \tjudgP{\tstruct{S}&}{\pfld_{x_1}(\hole_1)}{\tmapping{\btype'}}{w_1 x_1} \label{eq:prf-struct-holes-fld-h1} \\
    \tjudgP{\tstruct{S}&}{\hole_1}{\tmapping{\tstruct{S}}}{w_1}
\end{align}
By inversion on Eq.~\eqref{eq:prf-struct-holes-fld-h1}, $\hole_1 \neq \holet$.
Thus, by the inductive hypothesis,
\begin{align}
    \tjudgP{\btype_{h_1}&}{\hole_1'}{\tmapping{\tstruct{S}}}{w_1'} \\
    w_1 &= xw_1' \\
    \defns(S, x) &= \btype_{h_1} \\
    \hole_1'(\holet[.x]) &= \hole_1
\end{align}
Choose
\begin{align}
    \hole' &= \pfld_{x_1}(\hole_1') \\
    \btype_h &= \btype_{h_1} \\
    w' &= w_1'x_1
\end{align}
The required result follows immediately by applying {\sc TH-Fld}:
\begin{align}
    \hole'(\holet[.x]) &= \pfld_{x_1}(\hole_1'(\holet[.x])) = \pfld_{x_1}(\hole_1) = \hole \\
    w &= w_1x = xw_1'x = xw' \\
    \tjudgP{\btype_{h}&}{\pfld_{x_1}(\hole_1')}{\tmapping{\btype'}}{w_1'x_1}
\end{align}

\item Case {\sc TH-Sum, TH-Flatten}: Similar.

\item Case {\sc TH-FldAcc}: Then
\begin{align}
    H &= \hole_1[.x_1] \\
    w &= w_1 x_1 \\
    \tjudgP{\tstruct{S}&}{\hole_1[.x_1]}{\btype}{w_1 x_1} \\
    \tjudgP{\tstruct{S}&}{\hole_1}{\tstruct{S}}{w_1}
    \label{eq:prf-struct-holes-fldacc-h1} \\
    \defns(S, .x_1) &= \btype
\end{align}
We split on the cases $\hole_1 = \holet$ and $\hole_1 \neq \holet$.
In the former case, $w_1 = \epsilon$ by inversion on Eq.~\eqref{eq:prf-struct-holes-fldacc-h1};
the required result immediately follows by setting $\hole' = \holet$, $\btype_h = \btype$, $w' = \epsilon$, and $x = x_1$.
In the latter case, the proof proceeds similarly to the case for {\sc TH-Fld}.

\end{itemize}
\end{proof}

\subsection{Relating symbolic sum functions to their concrete counterparts}

The following lemma shows that the $\ConstSum$ function is well-defined with respect to the aggregation function templates synthesized for the \textsc{TE-MapInd} and \textsc{TE-MapUpd} rules. Furthermore, it relates the symbolic value of an aggregation of mapping (i.e., some application of a $\psum$ function) to the concrete value of the aggregation.

\begin{lemma}[Correctness of template synthesis for Sum]
  \label{lem:sum-correctness}
  Suppose
  \begin{align}
    (\hole_1, \hole_2, w) &\in \Hfam(\tint, \btype) \\
    \hole_1 &= \psum(\hole_1')  \\
    v &= \emapping{\btype}{n_1 \mapsto v_1, \dots, n_k \mapsto v_k} \\
    \forall v_i, \quad &\tjudg{\Env;\Guards}{v_i}{\btype} \label{eq:sum-correctness-vtype}
  \end{align}
  and let $\phi$ be the additional clauses generated when encoding $v$ into SMT. Then $\ConstSum(w, v)$ is well-defined, and for all $v_i$, the formula
  \begin{equation}
      \phi \implies
      \Encode(\hole_2(v_i) = \ConstSum(w, v_i))
  \end{equation}
  is valid.
\end{lemma}

\begin{example}
Consider the mapping
\[
\begin{array}{l}
v = \emapping{\tstruct{S}}{ \\
\qquad 1 \mapsto \estruct{S}{a \mapsto \emapping{\tint}{5 \mapsto 11}}, \\
\qquad 2 \mapsto \estruct{S}{a \mapsto \emapping{\tint}{4 \mapsto 6, 10 \mapsto 3}} \\
}
\end{array}
\]
It is only possible to sum over the nested mapping values at the $a$ field in each struct, so the templates corresponding to this mapping are
\begin{align*}
    \hole_1 &= \psum(\pflat(\pfld_a(\holet))) \\
    \hole_2 &= \psum(\holet[.a]) \\
    \Hfam(\tint, \tstruct{S}) &= \{(\hole_1, \hole_2, a)\}
\end{align*}
Let us write $v_2$ to be the struct value at index 2 of $v$.
The concrete aggregation values of $v$ and $v_2$ are given, respectively, by
\begin{align*}
    \ConstSum(a, v) &= 11 + 6 + 3 = 20 \\
    \ConstSum(a, v_2) &= 6 + 3 = 9
\end{align*}
Intuitively, the value of $\psum(v_2[.a])$ should be equal to the concrete sum $\ConstSum(a, v_2)$, which is what the lemma says:
\[
    \hole_2(v_2) = \psum(v_2[.a]) = 9
\]
This concludes the example of Lemma~\ref{lem:sum-correctness}.
\end{example}

\begin{proof}[Proof of Lemma~\ref{lem:sum-correctness}] By induction on $\btype$.
\begin{itemize}
\item Case $\btype = \tint$:
By inversion of the derivation of the synthesis judgment used for $\hole_2$, we see that the derivation must have used \textsc{TH-Hole}, so
\begin{align}
    \hole_2(v_i) &= v_i \\
    w &= \epsilon
\end{align}
By Lemma~\ref{lem:canonical-forms} and Eq.~\eqref{eq:sum-correctness-vtype}, we must have $v_i \in \mathbb{N}$. Thus, we have
\[
\ConstSum(w, v) = \sum_{j=1}^k v_j
\]
The required result follows immediately.

\item Case $\btype = \tmapping{\btype'}$:
By definition of $\Hfam$, we have
\begin{equation}
    \tjudgP{\tmapping{\btype'}}{\hole_2}{\tint}{w} \label{eq:prf-sum-correct-map-h2}
\end{equation}
By inversion on Eq.~\eqref{eq:prf-sum-correct-map-h2}, we see that \textsc{TH-Sum} must have been used.
\begin{align}
    \hole_2 &= \psum(\hole_2') \\
    \tjudgP{\tmapping{\btype'}&}{\hole_2'}{\tmapping{\tint}}{w}
\end{align}
By Eq.~\eqref{eq:sum-correctness-vtype} and Lemma~\ref{lem:canonical-forms},
\begin{align}
v_i &= \emapping{\btype'}{n_1 \mapsto v_1', \dots, n_q \mapsto v_q'}
\end{align}
Note that there must exist at least one $\hole_2''$ such that $(\hole_2, \hole_2'', w) \in \Hfam(\tint, \btype')$.
By applying the inductive hypothesis, we see that for every $p = 1 \dots q$, $\ConstSum(w, v_p')$ is well-defined. Thus, we have
\begin{equation}
    \ConstSum(w, v_i) = \sum_{p=1}^q \ConstSum(w, v_p')
\end{equation}
So $\ConstSum(w, v)$ is well-defined as well.
It follows that the following formula appears in the type of $v_i$:
\begin{align}
    \hole_2(\nu) &= \sum_{p=1}^q \ConstSum(w, v_p')
\end{align}
The required result follows by transitivity of equality.

\item Case $\btype = \tstruct{S}$:
By Lemma~\ref{lem:canonical-forms} and inversion on the derivation of Eq.~\eqref{eq:sum-correctness-vtype}, we must have
\begin{align}
v_i &= \estruct{S}{\dots, x \mapsto v_i', \dots} \\
\defns(S, x) &= \btype' \\
\tjudg{\Env;\Guards&}{v_i'}{\btype'} \label{eq:prf-sum-correct-vip-ty}
\end{align}
By definition of $\Hfam$, we have
\begin{equation}
    \tjudgP{\tstruct{S}}{\hole_2}{\tint}{w}
    \label{eq:prf-sum-correct-fld-h2}
\end{equation}
By inversion on the derivation of Eq.~\eqref{eq:prf-sum-correct-fld-h2}, $\hole_2 \neq \hole_t$. Thus, by Lemma~\ref{lem:struct-holes}, there exist $\btype', \hole_2', x, w'$ such that
\begin{align}
    \tjudgP{\btype'&}{\hole_2'}{\tint}{w'} \\
    w &= xw' \\
    \defns(S, x) &= \btype' \\
    \hole_2'(\holet[.x]) &= \hole_2 \label{eq:prf-sum-correct-fld-h2p}
\end{align}

Now, we construct a mapping constant $v'$ as follows:
\begin{equation}
    v' = \emapping{\btype'}{n_1 \mapsto v_1', \dots, n_k \mapsto v_k'}
\end{equation}
Observe that there exists some hole $\hole_2''$ such that $(\psum(\hole_2''), \hole_2', w') \in \Hfam(\tint, \btype')$. Thus, the conditions required for the inductive hypothesis are satisfied, so we conclude that $\ConstSum(w', v')$ is well-defined. From the definition of $\ConstSum$, we then have
\begin{align}
    \ConstSum(w', v') &= \sum_{j=1}^k \ConstSum(w', v_j') \\
    \ConstSum(w, v_i) &= \ConstSum(w', v_i')
    \label{eq:prf-sum-correct-fld-csvi} \\
    \ConstSum(w, v) &= \sum_{j=1}^k \ConstSum(w, v_j)
\end{align}
so $\ConstSum(w, v)$ is well-defined.
By Eq.~\eqref{eq:prf-sum-correct-fld-csvi} and the inductive hypothesis, $\hole_2'(v_i') = \ConstSum(w, v_i)$.
But then $v_i[.x] = v_i'$ is on the LHS of the implication when encoding to SMT, so the required result follows immediately since $\hole_2'(v_i[.x]]) = \hole_2(v_i)$ by Eq.~\eqref{eq:prf-sum-correct-fld-h2p}.

\end{itemize}

\end{proof}

\subsection{Progress and Preservation}

\begin{theorem}[Progress for expressions]
  \label{thm:progress-expr}
  If $\tjudg{\PrfCtx}{\expr}{\type}$, then $\expr$ is a value or there exists an $\expr'$ such that $\stepE{\expr}{\expr'}$.
\end{theorem}

\begin{proof} By induction on the derivation of $\tjudg{\PrfCtx}{\expr}{\type}$.

\begin{itemize}
\item Case {\sc TE-Nat, TE-True, TE-False, TE-StructConst, TE-MappingConst}: Then $\expr$ is a value.

\item Case {\sc TE-Sub}: Then
\begin{align}
    \type = \type' \\
    \tjudg{\PrfCtx}{\expr}{\type'} \\
    \tjudg{\PrfCtx}{\expr}{\type''} \label{eq:prf-prog-sub} \\
    \subty{\PrfCtx}{\type''}{\type'}
\end{align}
By applying the inductive hypothesis to Eq.~\eqref{eq:prf-prog-sub}, $\expr$ is a value or there exists an $e'$ such that $\stepE{\expr}{\expr'}$, which is exactly the required result.

\item Case {\sc TE-Var}: This case is impossible since the local environment is empty.

\item Case {\sc TE-Plus}: Then
\begin{align}
    \type &= \rtype{\tint}{\nu = \expr_1 + \expr_2} \\
    \expr &= \expr_1 + \expr_2 \\
    \tjudg{\PrfCtx&}{\expr_1}{\tint} \label{eq:prf-prog-plus-ih1} \\
    \tjudg{\PrfCtx&}{\expr_2}{\rtype{\tint}{\nu + \expr_1 \leq \pmaxint}} \label{eq:prf-prog-plus-ih2}
\end{align}
By applying the inductive hypothesis to Eq.~\eqref{eq:prf-prog-plus-ih1} (resp. Eq.~\eqref{eq:prf-prog-plus-ih2}), either $\expr_1$ is a value or $\stepE{\expr_1}{\expr_1'}$ (resp. $\expr_2$ and $\expr_2'$). By case analysis on these newly introduced facts:
\begin{itemize}
\item Case $\stepE{\expr_1}{\expr_1'}$: Apply {\sc EE-Ctx} so that $e' = e_1' + e_2$.
\item Case $\expr_1$ is a value and $\stepE{\expr_2}{\expr_2'}$: Apply {\sc EE-Ctx} so that $e' = e_1 + e_2'$.
\item Case $\expr_1, \expr_2$ are values: By Lemma~\ref{lem:canonical-forms}, Eq.~\eqref{eq:prf-prog-plus-ih1}, and Eq.~\eqref{eq:prf-prog-plus-ih2},
\begin{align}
    \expr_1 &= n_1 \\
    \expr_2 &= n_2
\end{align}
Apply {\sc EE-Plus} so that $e' = n_1 + n_2$.
\end{itemize}

\item Case {\sc TE-Minus}, {\sc TE-Rel}: Similar.

\item Case {\sc TE-MapInd}: Then
\begin{align}
    \expr &= \eind{\expr_1}{\expr_2} \\
    &\tjudg{\PrfCtx}{\expr_1}{\tmapping{\btype}} \label{eq:prf-prog-mind-t1} \\
    &\tjudg{\PrfCtx}{\expr_2}{\tint} \label{eq:prf-prog-mind-t2}
\end{align}
By applying the inductive hypothesis to Eq.~\eqref{eq:prf-prog-mind-t1} (resp. Eq.~\eqref{eq:prf-prog-mind-t2}), either $\expr_1$ is a value or $\stepE{\expr_1}{\expr_1'}$ (resp. $\expr_2$ and $\expr_2'$). By case analysis on these newly introduced facts:
\begin{itemize}
\item Case $\stepE{\expr_1}{\expr_1'}$: Apply {\sc EE-Ctx} so that $e' = \eind{e_1'}{e_2}$.
\item Case $\expr_1$ is a value and $\stepE{\expr_2}{\expr_2'}$: Apply {\sc EE-Ctx} so that $e' = \eind{e_1}{e_2'}$.
\item Case $\expr_1, \expr_2$ are values: By Lemma~\ref{lem:canonical-forms}, Eq.~\eqref{eq:prf-prog-mind-t1}, and Eq.~\eqref{eq:prf-prog-mind-t2},
It follows that
\begin{align}
    \expr_1 &= \emapping{\btype}{n_1 \mapsto v_1, \dots, n_k \mapsto v_k} \\
    \expr_2 &= n
\end{align}
If there exists $i \in \{1, \dots, k\}$ such that $n = n_i$, then apply {\sc EE-MapInd1} so that $e' = v_i$.
Otherwise, apply {\sc EE-MapInd2} so that $e' = \ZeroVal{\btype}$.
\end{itemize}

\item Case {\sc TE-MapUpd}: Similar.

\item Case {\sc TE-SctInd}: Then
\begin{align}
    \expr &= \eind{\expr_1}{.x} \\
    &\tjudg{\PrfCtx}{\expr_1}{\tstruct{S}} \label{eq:prf-prog-sind-t1} \\
    \defns(S, x) &= \btype
\end{align}
By applying the inductive hypothesis to Eq.~\eqref{eq:prf-prog-sind-t1}, either $\expr_1$ is a value or $\stepE{\expr_1}{\expr_1'}$. By case analysis on these newly introduced facts:
\begin{itemize}
\item Case $\stepE{\expr_1}{\expr_1'}$: Apply {\sc EE-Ctx} so that $e' = \eind{e_1'}{e_2}$.
\item Case $\expr_1$ is a value: By Lemma~\ref{lem:canonical-forms} and Eq.~\eqref{eq:prf-prog-sind-t1},
\begin{align}
    \expr_1 &= \estruct{S}{x_1 \mapsto v_1, \dots, x_n \mapsto v_n} \\
    \defns(S) &= \{x_1:\btype_1, \dots, x_n:\btype_n\}
\end{align}
It follows that $x = x_i$ for some $x_i \in \{x_1, \dots, x_n\}$, so
apply {\sc TE-SctInd} so that $e' = v_i$.
\end{itemize}

\item Cases {\sc TE-MapUpd}, {\sc TE-SctUpd}: Similar.

\end{itemize}
\end{proof}

\begin{theorem}[Preservation for expressions]
  \label{thm:pres-expr}
  If $\tjudg{\PrfCtx}{\expr}{\type}$ and $\stepE{\expr}{\expr'}$, then $\tjudg{\PrfCtx}{\expr'}{\type}$.
\end{theorem}

\begin{proof} By induction on the derivation of $\stepE{\expr}{\expr'}$.

\begin{itemize}
\item Case {\sc EE-Ctx}: Then
\begin{align}
    \expr &= \Ectx(\expr_1) \\
    \stepE{\expr_1&}{\expr_1'} \label{eq:prf-pres-ctx-step} \\
    \expr' &= \Ectx(\expr_1')
\end{align}
By Lemma~\ref{lem:eval-ctx-arg-typed}, there exists a $\type_1$ such that
\begin{equation}
    \tjudg{\PrfCtx}{\expr_1}{\type_1} \label{eq:prf-pres-ctx-e1-type}
\end{equation}
By applying the inductive hypothesis to Eq.~\eqref{eq:prf-pres-ctx-step} and Eq.~\eqref{eq:prf-pres-ctx-e1-type},
\begin{equation}
    \tjudg{\PrfCtx}{\expr_1'}{\type_1}
\end{equation}
The required result is then obtained by applying Lemma~\ref{lem:eval-ctx-swap-typed}.

\item Case {\sc EE-Plus}: Then
\begin{align}
    \expr &= n_1 + n_2 \\
    \expr' &= n_1 +_\mathbb{N} n_2 \\
    \type &= \rtype{\tint}{\nu = n_1 + n_2}
\end{align}
By inversion on the derivation of $\tjudg{\PrfCtx}{\expr}{\type}$, we see that the derivation must have used \textsc{TE-Plus}:
\begin{align}
    &\tjudg{\PrfCtx}{n_1}{\tint} \\
    &\tjudg{\PrfCtx}{n_2}{\rtype{\tint}{\nu + n_1 \leq \pmaxint}}
\end{align}
Since 
\begin{equation}
    \expr = n_1 + n_2 \land n_1 + n_2 \leq \pmaxint \implies \expr \leq \pmaxint
\end{equation}
is valid and $\expr' = (n_1 +_\mathbb{N} n_2)$, we can apply \textsc{TE-Nat} to obtain
\begin{equation}
    \tjudg{\Env;\Guards}{\expr'}{\rtype{\tint}{\nu = n_1 +_\mathbb{N} n_2}}
\end{equation}
Applying \textsc{Sub-Base} to the above typing judgment then yields the desired result.

\item Case {\sc EE-Minus}, {\sc EE-Rel}: Similar.

\item Case {\sc EE-MapInd1}: Then
\begin{align}
    \expr &= \expr_1[\expr_2] \\
    \expr_1 &= \emapping{\btype}{n_1 \mapsto \val_1, \dots, n_i \mapsto \val_i, \dots, n_k \mapsto \val_k} \\
    \expr_2 &= n_i \\
    \expr' &= v_i
\end{align}
By inversion on the derivation of $\tjudg{\Env;\Guards}{\expr}{\type}$, we see that the derivation must have used \textsc{T-MapInd}:
\begin{align}
    \type &= \rtype{\btype}{\nu = \eind{\expr_1}{\expr_2} \land \phi} \\
    &\tjudg{\PrfCtx}{\expr_1}{\tmapping{\btype}} \\
    &\tjudg{\PrfCtx}{\expr_2}{\tint} \\
    \Hfam(\tint, \btype) &= \{(\hole_{11}, \hole_{12}, w_1), \dots, (\hole_{\ell 1}, \hole_{\ell 2}, w_\ell)\} \\
    \phi &= \bigwedge_{j=1}^\ell \hole_{j2}(\nu) \leq \hole_{j1}(\expr_1)
\end{align}
By inversion on the derivation of $\tjudg{\PrfCtx}{\expr_1}{\tmapping{\btype}}$, we see that the derivation must have used \textsc{Sub-Base} and \textsc{T-MapConst}:
\begin{align}
    &\subty{\PrfCtx}{\tau'}{\tmapping{\btype}} \\
    &\tjudg{\PrfCtx}{\expr_1}{\tau'} \\
    \tau' &= \rtype{\tmapping{\btype}}{\nu = e_1 \land \phi'} \\
    \mathsf{Encode}&\left(\left(\nu = e_1 \land \phi'\right)[\subst{\nu}{\expr_1}]\right) \implies \etrue \quad\text{valid} \\
    \text{for $m = 1 \dots k$,}\qquad &\tjudg{\PrfCtx}{\val_m}{\btype} \label{eq:prf-pres-mapind-v} \\
    \phi' &= \bigwedge_{j=1}^\ell \left(\hole_{j1}(\nu) = \ConstAgg(\btype, \hole_{j1}, w_j, e_1)\right)
\end{align}
By Lemma~\ref{lem:sum-correctness} and the definition of $\ConstSum$, we have
\begin{equation}
    \ConstSum(w_j, e_1) \geq \ConstSum(w_j, v_i)
\end{equation}
Thus, the following implication is valid:
\[
\hole_{j1}(\nu) = \ConstSum(w_j, e_1) \land \hole_{j2}(v_i) = \ConstSum(w_j, v_i) \implies \hole_{j1}(\nu) \geq \hole_{j2}(v_i)
\]
Consequently, the implication
\begin{equation}
  \mathsf{Encode}(\nu = v_i)
  \implies
  \mathsf{Encode}(\nu = \eind{\expr_1}{n_i} \land \phi)
\end{equation}
is also valid, so the required result is then obtained by applying \textsc{Sub-Base} to Eq.~\eqref{eq:prf-pres-mapind-v}.

\item Case {\sc EE-MapInd2}: Similar.

\item Case {\sc EE-MapUpd1}: Then
\begin{align}
    \expr &= \eupd{\expr_1}{\expr_2}{\expr_3} \\
    \expr_1 &= \emapping{\btype}{n_1 \mapsto \val_1, \dots, n_i \mapsto \val_i, \dots, n_k \mapsto \val_k} \\
    \expr_2 &= n_i \\
    \expr_3 &= \val \\
    \expr' &= \emapping{\btype}{n_1 \mapsto \val_1, \dots, n_i \mapsto \val, \dots, n_k \mapsto \val_k}
\end{align}
By inversion on the derivation of $\tjudg{\PrfCtx}{\expr}{\type}$, we see that the derivation must have used \textsc{T-MapUpd}:
\begin{align}
    \type &= \rtype{\btype}{\nu = \eupd{\expr_1}{n_i}{\val} \land \phi} \\
    &\tjudg{\PrfCtx}{\expr_1}{\tmapping{\btype}} \\
    &\tjudg{\PrfCtx}{n_i}{\tint} \\
    &\tjudg{\PrfCtx}{v}{\btype} \\
    \Hfam(\tint, \btype) &= \{(\hole_{11}, \hole_{12}, w_1), \dots, (\hole_{\ell 1}, \hole_{\ell 2, w_\ell})\} \\
    \phi &= \bigwedge_{j=1}^\ell \hole_{j1}(\nu) = \hole_{j1}(\expr_1) - \hole_{j2}(\expr_1[n_i]) + \hole_{j2}(\val)
\end{align}
By inversion on the derivation of $\tjudg{\PrfCtx}{\expr_1}{\tmapping{\btype}}$, we see that the derivation must have used \textsc{Sub-Base} and \textsc{T-MapConst}:
\begin{align}
    &\subty{\PrfCtx}{\tau'}{\tmapping{\btype}} \\
    &\tjudg{\PrfCtx}{\expr_1}{\tau'} \\
    \tau' &= \rtype{\tmapping{\btype}}{\nu = e_1 \land \phi'} \\
    &\mathsf{Encode}\left(\left(\nu = e_1 \land \phi'\right)[\subst{\nu}{\expr_1}]\right) \implies \etrue \quad\text{valid} \\
    \text{for $m = 1 \dots k$,}\qquad &\tjudg{\PrfCtx}{\val_m}{\btype} \label{eq:prf-pres-mapupd-v} \\
    \phi' &= \bigwedge_{j=1}^\ell \left(\hole_{j1}(\nu) = \ConstAgg(\btype, \hole_{j1}, w_j, e_1)\right)
\end{align}
By definition of $\ConstAgg$, the following equation holds:
\begin{align}
    \ConstSum(w_j, e') - \ConstSum(w_j, e_1) &= \ConstSum(w_j, v) - \ConstSum(w_j, v_i)
\end{align}
Together, the above equation and Lemma~\ref{lem:sum-correctness} imply that the following implication is valid:
\begin{equation}
\begin{array}{rlr}
    \hole_{j1}(\nu) &= \ConstSum(w_j, e')
    &\land \\
    \hole_{j1}(e_1) &= \ConstSum(w_j, e_1)
    &\land \\
    \hole_{j2}(v) &= \ConstSum(w_j, v)
    &\land \\
    \hole_{j2}(v_i) &= \ConstSum(w_j, v_i) &\\
    \implies \\
    \multicolumn{3}{c}{\hole_{j1}(\nu) = \hole_{j1}(e_1) - \hole_{j2}(v) + \hole_{j2}(v_i)}
\end{array}
\end{equation}

Therefore we can construct a $\phi''$ defined as
\begin{align}
    \phi'' &= \bigwedge_{j=1}^\ell \left(\hole_{j1}(e') = \ConstAgg(T, \hole_{j1}, w_j, e')\right)
\end{align}
The required result is then obtained by applying \textsc{TE-MappingConst} and \textsc{Sub-Base}.

\item Case {\sc EE-MapUpd2}: Similar.

\item Case {\sc EE-SctInd}: Similar to \textsc{EE-MapInd1}.

\item Case {\sc EE-SctUpd}: Similar to \textsc{EE-MapUpd1}.
\end{itemize}

\end{proof}

\section{Statement Rules}
\label{sec:appendix-stmt-rules}

\subsection{Syntax and Judgments}

\[
\begin{array}{r c l l}
\Sctx & ::= & & \textbf{Statement evaluation context} \\
&& \odot \mid \slet{x}{\type}{\Ectx} \mid \sassert \Ectx \mid \sassume \Ectx & \\
&\mid& \sif{\Ectx}{s_1}{s_2}{j} & \\
&\mid& \scommit{v_1 \sto x_1, v_2 \sto x_2, \dots, \Ectx \sto x_i, e_{i+1} \sto x_{i+1}, \dots} & \\
&\mid& \scall{x:\type}{f}{v_1, v_2, \dots, v_{i-1}, \Ectx, e_{i+1}, \dots} & \\
\\
\store & \in & Var \to Value & \textbf{State variable store} \\
\\
\defnsE & ::= & \epsilon \mid (\declFun{f}{x_1:\ \type_1, \dots, x_n:\ \type_n}{\type_r}{s}{e}), \defnsE & \textbf{Global declarations for evaluation} \\
\\
\multicolumn{3}{l}{\stepS{s}{\store}{s}{\store'}{\stmtSub}} & \textbf{Statement evaluation relation} \\
\\
\stmtSub & ::= & (x_1 \mapsto e_1, \dots, x_n \mapsto e_n) & \textbf{Local variable substitution}
\end{array}
\]

Similar to the expression evaluation context, the statement evaluation context $\Sctx$ defines the evaluation order of expressions nested in statements.
The state variable store $\store$ maps each state variable to its concrete value.
The statement evaluation relation $\stepS{s}{\store}{s}{\store'}{\stmtSub}$ asserts that starting from state variable store $\store$, $s$ will step to $s'$, update the store to $\store'$, and generate a local variable bindings $\stmtSub$ to be applied as a substitution to the next statement.
We will assume that $\defnsE$ is fixed and implicitly defined everywhere.

We also treat the local variable bindings $\stmtSub$ as a substitution: $\stmtSub(t)$, $\stmtSub(\type)$, and $\stmtSub(\Guards)$ are defined as substitution over $t$, substitution over $\type$, and pointwise substitution over the terms in $\Guards$, respectively.
We define the substitution of an environment $\stmtSub(\Env)$ recursively as follows:
\begin{align*}
    \stmtSub(\epsilon) &= \epsilon \\
    \stmtSub(x: \type, \Env) &= \begin{cases}
        \stmtSub(\Env) & \text{ if } x \in \dom(\Env) \\
        x: \stmtSub(\type), \stmtSub(\Env) & \text{ if } x \notin \dom(\Env)
    \end{cases}
\end{align*}
Intuitively, the effect on the environment is that all variables in $\stmtSub$ are dropped, and that all other variables have the substitution applied to their types.

\subsection{Statement evaluation rules}

\begin{figure}
    \centering
\begin{mathpar}
\inferrule*[right=ES-Ctx]{
  \stepE{\expr}{\expr'}
}{
  \stepS{\Sctx(\expr)}{\store}{\Sctx(\expr')}{\store}{\epsilon}
}
\\
\inferrule*[right=ES-AssertTrue]{ }{
  \stepS{\sassert \etrue}{\store}{\sskip}{\store}{\epsilon}
}
\and
\inferrule*[right=ES-AssumeTrue]{ }{
  \stepS{\sassume \etrue}{\store}{\sskip}{\store}{\epsilon}
}
\\
\inferrule*[right=ES-Seq1]{
  \stepS{s_1}{\store}{s_1'}{\store'}{\stmtSub} \\
  \stmtSub = \overrightarrow{x_i \mapsto v_i}
}{
  \stepS{s_1;s_2}{\store}{s_1';\stmtSub(s_2)}{\store'}{\stmtSub}
}
\and
\inferrule*[right=ES-Seq2]{ }{
  \stepS{\sskip;s}{\store}{s_2}{\store}{\epsilon}
}
\and
\inferrule*[right=ES-Seq3]{ }{
  \stepS{\sassume \efalse;s_2}{\store}{\sassume \efalse}{\store}{\epsilon}
}
\\
\inferrule*[right=ES-Let]{ }{
  \stepS{\slet{x}{\type}{\val}}{\store}{\sskip}{\store}{(x \mapsto v)}
}
\\
\inferrule*[right=ES-Call]{
  \defnsE(f) = \declFun{f}{x_1:\ \type_1, \dots, x_n:\ \type_n}{\type_r}{s}{e} \\\\
  s' = s \text{ with variable declarations alpha renamed}
  \\
  \stmtSub = (\subst{x_1}{v_1}, \dots, \subst{x_n}{v_n})
}{
  \stepS{\scall{x: \type}{f}{v_1, \dots, v_n)}}{\store}
  {(\stmtSub(s);\slet{x}{\type}{e})}{\store}{\epsilon}
}
\\
\inferrule*[right=ES-Fetch]{
  \text{for $i = 1..n$,\quad} \sigma(x_i') = v_i
}{
  \stepS{\sfetch{x_1' \sas x_1, \dots, x_n' \sas x_n}}{\store}{\sskip}{\store}{(x_1 \mapsto v_1, \dots, x_n \mapsto v_n)}
}
\\
\inferrule*[right=ES-Commit]{
  \dom(\store) = \{x_1, \dots, x_n\}
}{
  \stepS{\scommit{v_1 \sto x_1, \dots, v_n \sto x_n}}{\store}{\sskip}{(x_1 \mapsto v_1, \dots, x_n \mapsto v_n)}{\epsilon}
}
\\
\inferrule*[right=ES-IfTrue]{
  j = x_1:\type_i = \phi(x_{11}, x_{12}),\dots,x_n:\type_n = \phi(x_{n1}, x_{n2}) \\
  s_j = \slet{x_1}{\type_1}{x_{11}};\cdots;\slet{x_n}{\type_n}{x_{n1}}
}{
  \stepS{\sif{\etrue}{s_1}{s_2}{j}}{\store}{s_1;s_j}{\store}{\epsilon}
}
\\
\inferrule*[right=ES-IfFalse]{
  j = x_1:\type_i = \phi(x_{11}, x_{12}),\dots,x_n:\type_n = \phi(x_{n1}, x_{n2}) \\
  s_j = \slet{x_1}{\type_1}{x_{12}};\cdots;\slet{x_n}{\type_n}{x_{n2}}
}{
  \stepS{\sif{\efalse}{s_1}{s_2}{j}}{\store}{s_2;s_j}{\store}{\epsilon}
}
\end{mathpar}
    \caption{Statement evaluation rules}
    \label{fig:stmt-eval-rules}
\end{figure}

The statement evaluation rules are shown in Figure~\ref{fig:stmt-eval-rules}.

The {\sc ES-Ctx} rule steps an expression contained inside of a statement.

The {\sc ES-Seq1}, {\sc ES-Seq2}, and {\sc ES-Seq3} rules define how statements may be composed.
If the first statement in the sequence can step, then the {\sc ES-Seq1} will step it.
If the first statement in the sequence happens to be $\sskip$, the {\sc ES-Seq2} rule advances to the next second statement in the sequence.
Lastly, if the head of the sequence contains a failing runtime check $\sassume \efalse$, then the {\sc ES-Seq3} rule propagates the failed runtime check.

Assertions (static checks) and assumptions (dynamic checks) are respectively handled by the {\sc ES-AssertTrue} and {\sc ES-AssumeTrue} rules. No equivalent rule is defined for $\efalse$ for either, as we will later prove that 1) the type system ensures that a well-typed statement cannot step to $\sassert{\efalse}$; and 2) $\sassume{\efalse}$ corresponds to a failed runtime check, at which point execution halts.

The {\sc ES-Fetch} and {\sc ES-Commit} rules encode the simultaneous load/store, as described in the main text.

\subsection{Typing Rules Omitted in the Main Text}

We omitted the rule for $\sskip$ in the main text, so we provide it here.

\begin{mathpar}
\inferrule*[Right=TS-Skip]{ }{
  \tjudgS{\Env;\Lock,\Guards}{\sskip}{\Env;\Lock,\Guards}
}
\end{mathpar}

\subsection{Store and Globals Typing}

The \emph{store typing judgment} $\tjudgStore{\store}$ means that all state variables are present in $\store$ and that the state variables are well-typed with respect to "each other". That is, the refinements of the state variables may only have the other state variables as free variables.

\newcommand{\tjudgGlobals}[1]{\vdash #1}

The \emph{global declarations typing judgment}, which is of the form $\tjudgGlobals{\defnsE}$, means that all of the functions in the global declarations structure $\defnsE$ are well-typed.

\begin{mathpar}
\inferrule*[right=T-Store]{
  \store = (x_1 \mapsto v_1, \dots, x_n \mapsto v_n) \\\\
  \text{for $i = 1 \dots n,$}
  \and
  \defns(x_i) = \type_i
  \and
  \tjudg{\Env;\epsilon}{v_i}{\type_i} \\\\
  \Env = x_1 : \type_1, \dots, x_n : \type_n
}{
  \tjudgStore{\store}
}
\\
\inferrule*[right=T-GloDecls]{
  \defnsE = decl_1, \dots, decl_n \\\\
  \text{for }i = 1 \dots n, \and
  decl_i = \declFun{f_i}{\overrightarrow{x_{ij}:\ \type_{ij}}}{\type_{r_i}}{s_i}{e_i} \and
  \vdash decl_i
}{
  \tjudgGlobals{\defnsE}
}
\end{mathpar}

\section{Progress and Preservation for Statements}
\label{sec:appendix-stmt-proofs}

Similar to progress for expressions, progress for statements is straightforwardly proven.
In contrast, progress for statements is harder to prove as substitution can change the resulting environment and guard predicates.

\subsection{Progress}

To ease the notational burden, let us first assume that we use a single global declarations $\defnsE$ and that $\defnsE$ is well-typed.

\begin{theorem}[Progress for statements]
  If $\tjudgS{\PrfCtxPg}{s}{\Env;L',\Guards}$ and $\tjudgStore{\store}$, then $s = \sskip$, or $s = \sassume \efalse$, or there exist $s', \store', \store_{sub}$ such that $\stepS{s}{\store}{s'}{\store'}{\store_{sub}}$.
\end{theorem}
\begin{proof} By induction on the derivation of $\tjudgS{\PrfCtxPg}{s}{\Env;L',\Guards}$.

\begin{itemize}

\item Case {\sc TS-Skip}: Follows immediately.

\item Case {\sc TS-Seq}: Then
\begin{align}
    s &= s_1;s_2 \\
    \tjudgS{\PrfCtxPg&}{s_1}{\Env;L'',\Guards'} \label{eq:prf-sprog-seq-s1}
\end{align}
By applying the inductive hypothesis to Eq.~\eqref{eq:prf-sprog-seq-s1}, either $s_1 = \sskip$ or $s_1 = \sassume \efalse$ or there exists $s_1',\store, \store_{sub}$ such that $\stepS{s_1}{\store}{s_1'}{\store}{\store_{sub}}$.
The required result then follows immediately from an application of {\sc ES-Seq2}, {\sc ES-Seq3}, or {\sc ES-Seq1}, respectively.

\item Case {\sc TS-Let}: Then
\begin{align}
    s &= \slet{x}{\type}{\expr}
\end{align}
By Theorem~\ref{thm:progress-expr}, either $\expr$ is a value or there exists $\expr'$ such that $\stepE{\expr}{\expr'}$.
The required result then follows immediately from an application of {\sc ES-Let} or {\sc ES-Ctx}, respectively.

\item Case {\sc TS-Assert}: Then
\begin{align}
    s &= \sassert e \\
    \tjudg{\PrfCtx&}{\expr}{\rtype{\tbool}{\nu = \etrue}} \label{eq:prf-sprog-assert-e}
\end{align}
By Theorem~\ref{thm:progress-expr}, either $\expr$ is a value or there exists an $\expr'$ such that $\stepE{\expr}{\expr'}$.
In the former case, we must have $e = \etrue$ due to Eq.~\eqref{eq:prf-sprog-assert-e} and Lemma~\ref{lem:canonical-bool-value}, so the required result follows from an application of {\sc ES-AssertTrue}.
In the latter case, the required result follows from an application of {\sc ES-Ctx}.

\item Case {\sc TS-Assume}: Similar.

\item Case {\sc TS-Fetch}: Follows immediately from an application of {\sc ES-Fetch}.

\item Case {\sc TS-Commit}: Then
\begin{align}
    s &= \scommit{\expr_1 \sto x_1, \dots, \expr_n \to x_n}
\end{align}
By Theorem~\ref{thm:progress-expr}, every $\expr_i$ is a value or there exists an $\expr_i'$ such that $\stepE{\expr_i}{\expr_i'}$.
Let $1 \leq k \leq i$ be the first index where $e_k$ is not a value.
If $k$ does not exist, then all $e_i$'s are values, so the required result is obtained by applying {\sc ES-Commit}.
Otherwise, there exists an $\expr_k'$ such that $\stepE{\expr_k}{\expr_k'}$, so the required result is obtained by applying {\sc ES-Ctx}.

\item Case {\sc TS-Call}: Similar.

\item Case {\sc TS-If}: Then
\begin{align}
    s &= \sif{\expr}{s_1}{s_2}{j} \\
    \tjudg{\PrfCtx&}{\expr}{\tbool} \label{eq:prf-sprog-if-e}
\end{align}
By Theorem~\ref{thm:progress-expr}, either $\expr$ is a value or there exists an $\expr'$ such that $\stepE{\expr}{\expr'}$.
In the former case, Eq.~\eqref{eq:prf-sprog-if-e} and Lemma~\ref{lem:canonical-forms} implies that $e = \etrue$ or $e = \efalse$, so the required result is obtained by applying {\sc ES-IfTrue} or {\sc ES-IfFalse}, respectively.
In the latter case, the required result is obtained by applying {\sc ES-Ctx}.

\end{itemize}
\end{proof}

\subsection{Preservation}

\begin{theorem}[Preservation for Statements]
  \label{thm:pres-stmt-simple}
  If all of the following hold:
  \begin{enumerate}[label=(\alph*)]
  \item $\tjudgS{\PrfCtxS}{s}{\Env;L',\Guards}$
  \item $\tjudgStore{\store}$
  \item $\stepS{s}{\store}{s'}{\store'}{\store_{sub}}$
  \end{enumerate}
  then there exist $\Env', \Guards'$ such that
  \begin{enumerate}[label=(\roman*)]
  \item $\tjudgS{\epsilon;\PrfGdS'}{s'}{\Env';L',\Guards'}$
  \item $\tjudgStore{\store'}$
  \end{enumerate}
\end{theorem}

Intuitively, Theorem~\ref{thm:pres-stmt-simple} says that if $s$ is well-typed and can step to $s'$, then $s'$ is also well-typed with output environment $\Env'$ and guards $\Guards'$.

The overall strategy for proving Theorem~\ref{thm:pres-stmt-simple} is by induction on the derivation of clause (c).
Unfortunately, this strategy does not work well on Theorem~\ref{thm:pres-stmt-simple} as-is.
The primary difficulty lies in the {\sc TS-Seq1} case, as the environment and guard predicates obtained by applying the inductive hypothesis are completely arbitrary.
Thus, we instead prove a strengthened version of preservation (Theorem~\ref{thm:pres-stmt}) that then implies Theorem~\ref{thm:pres-stmt-simple}.
The following subsections outline the prerequisite lemmas and the required strengthening.

\subsection{Useful lemmas}

Since we must reason about how the environment and guard predicates change as a statement is evaluated, it is difficult to state the preservation theorem so that the inductive hypothesis will be strong enough to use in the proof.
Thus, we first present some lemmas that characterize properties of our type system; this will help motivate the correct strengthening of the preservation theorem.

The first of these lemmas formalizes the fact that no bindings or guards are deleted or modified when typing a statement.

\begin{lemma}
  \label{lem:persistent-bind}
  If $\tjudg{\Env_1;L,\Guards_1}{s}{\Env_2;L',\Guards_2}$, then there exists $\Env_2', \Guards_2'$ such that $\Env_2 = \Env_1, \Env_2'$ and $\Guards_2 = \Guards_1, \Guards_2'$.
\end{lemma}
\begin{proof} By induction on the derivation of $\tjudg{\Env_1;L,\Guards_1}{s}{\Env_2;L',\Guards_2}$.
\end{proof}

\subsubsection{Guard Lemmas}

The following two lemmas state that the predicate $\etrue$ can be safely removed from the guards.

\begin{lemma}
  \label{lem:contract-true-expr}
  If $\tjudg{\Env;\etrue,\Guards}{\expr}{\type}$, then $\tjudg{\Env;\Guards}{\expr}{\type}$.
\end{lemma}
\begin{proof} By induction on the derivation of $\tjudg{\Env;L,\etrue,\Guards}{\expr}{\type}$.
\end{proof}

\begin{lemma}
  \label{lem:contract-true-stmt}
  If $\tjudgS{\Env;L,\etrue,\Guards}{s}{\Env;L',\Guards'}$, then there exists $\Guards''$ such that $\Guards' = \etrue, \Guards''$ and $\tjudgS{\Env;L,\Guards}{s}{\Env;L',\Guards''}$.
\end{lemma}
\begin{proof} By induction on the derivation of
$\tjudgS{\Env;L,\etrue,\Guards}{s}{\Env;L',\Guards'}$.
\end{proof}

The following four lemmas encode the idea of precondition strengthening in the guards.
In particular, well-typedness is maintained when 1) adding an additional predicate to the guard; or 2) swapping a predicate with a stronger predicate.

\begin{lemma}
  If $\tjudg{\Env;\Guards}{\expr}{\type}$, then $\tjudg{\Env;\Guards, t}{\expr}{\type}$.
\end{lemma}
\begin{proof} By induction on the derivation of $\tjudg{\Env;\Guards}{\expr}{\type}$.
\end{proof}

\begin{lemma}
  If $\tjudg{\Env;t,\Guards}{\expr}{\type}$ and $\subty{\Env;\Guards}{\rtype{\tbool}{\nu = t'}}{\rtype{\tbool}{\nu = t}}$, then $\tjudg{\Env;t',\Guards}{\expr}{\type}$.
\end{lemma}
\begin{proof} Similar.
\end{proof}

\begin{lemma} \label{lem:weaken-stmt1}
  For all $t$, if $\tjudgS{\Env_1;\Guards_1}{s}{\Env_2;\Guards_2}$, then $\tjudgS{\Env_1;\Guards_1, t}{s}{\Env_2;\Guards_2,t}$.
\end{lemma}
\begin{proof} By induction on the derivation of $\tjudgS{\Env_1;\Guards_1}{s}{\Env_2;\Guards_2}$.
\end{proof}

\begin{lemma} \label{lem:weaken-stmt2}
  For all $t$, if $\tjudgS{\Env_1;t,\Guards_1}{s}{\Env_2;t,\Guards_2}$ and $\subty{\Env_1;\Guards_1}{\rtype{\tbool}{\nu = t'}}{\rtype{\tbool}{\nu = t}}$, then $\tjudgS{\Env_1;t',\Guards_1}{s}{\Env_2;t',\Guards_2}$.
\end{lemma}
\begin{proof} Similar.
\end{proof}

\subsubsection{Substitution Lemmas}

\begin{lemma}[Substitution for expressions]
  \label{lem:subst-expr}
  If all of the following are true:
  \begin{enumerate}[label=(\alph*)]
  \item $\stmtSub = (\subst{x_1}{\expr_1}, \dots, \subst{x_n}{\expr_n})$
  \item $\tjudg{x_1:\type_1, \dots, x_n: \type_n, \Env;\Guards}{\expr}{\type}$ \label{lem:subst-expr-etype}
  \item $\tjudg{\Env;\Guards}{\expr_i}{\type_i}$ for $i = 1\dots,n$
  \end{enumerate}
  then $\tjudg{\stmtSub(\Env);\stmtSub(\Guards)}{\stmtSub(e)}{\stmtSub(\type)}$
\end{lemma}
\begin{proof} By induction on the derivation of \ref{lem:subst-expr-etype}.
\end{proof}

\begin{lemma}[Substitution for statements]
  \label{lem:subst-stmt}
  If all of the following are true:
  \begin{enumerate}[label=(\alph*)]
  \item $\stmtSub = (\subst{x_1}{\expr_1}, \dots, \subst{x_n}{\expr_n})$
  \item $\tjudgS{x_1 : \type_1, \dots, x_n: \type_n, \Env_1;\Guards_1}{s}{\Env_2;\Guards_2}$ \label{lem:subst-stmt-stype}
  \item $\tjudg{\Env_1;\Guards_1}{\expr_i}{\type_i}$ for $i = 1\dots,n$,
  \end{enumerate}
  then $\tjudgS{\stmtSub(\Env_1);L_1,\stmtSub(\Guards_1)}{\stmtSub(s)}{\stmtSub(\Env_2);L_2,\stmtSub(\Guards_2)}$.
\end{lemma}
\begin{proof} By induction on the derivation of \ref{lem:subst-stmt-stype}.
\end{proof}

\subsubsection{Guard Change}

Let $A$, $D$ be guard predicates.
We define a {\it guard change relation} $\GuardChange{A}{D}{\Guards}{\Guards'}$ to mean that there exists some $\Guards''$ such that $\Guards = D,\Guards''$ and $\Guards' = A,\Guards''$.
Intuitively, this says that $\Guards'$ is obtained by removing the terms $D$ from $\Guards$ and then adding the terms $A$.
Alternatively, a guard change can be viewed as a partitioning of the terms in $\Guards$ and $\Guards'$ into three parts: terms exclusive to $\Guards$ (i.e., the terms in $D$), terms exclusive to $\Guards'$ (i.e., the terms in $A$), and terms shared by both (i.e., the terms in $\Guards''$).

\begin{example} The following is a valid guard change:
  \[
    \GuardChange{\{\etrue\}}{\{11 > 10\}}{11 > 10, x < 100}{\etrue, x < 100}
  \]
\end{example}

\subsubsection{Evaluation Context Lemmas}

\begin{lemma}[Inversion for statement eval context type]
  \label{lem:eval-sctx-arg-typed}
  If $\tjudgS{\Env_1;\Guards_1}{\Sctx(\expr)}{\Env_2;\Guards_2}$, then there exists a $\type$ such that $\tjudg{\Env_1;\Guards_1}{\expr}{\type}$.
\end{lemma}
\begin{proof} By induction on the derivation of $\tjudgS{\Env_1;\Guards_1}{\Sctx(\expr)}{\Env_2;\Guards_2}$.
\end{proof}

\begin{lemma}[Well-typedness of swapping statement eval context argument]
  \label{lem:eval-sctx-swap-typed}
  If $\tjudgS{\Env_1;\Guards_1}{\Sctx(\expr)}{\Env_2;\Guards_2}$ and $\tjudg{\Env_1;\Guards_1}{\expr}{\type}$ and $\tjudg{\Env_1;\Guards_1}{\expr'}{\type}$, then there exists a guard change $\GuardChange{A}{D}{\Guards_2}{\Guards_2'}$ such that $\tjudgS{\Env_1;\Guards_1}{\Sctx(\expr')}{\Env_2;\Guards_2'}$.
\end{lemma}
\begin{proof} By induction on $\Ectx_s$.
Most cases have $A = \epsilon$ and $D = \epsilon$.
The exceptions are the cases for $\sassert{\Ectx(e)}$ and $\sif{\Ectx(e)}{s_1}{s_2}$, where $A = \Ectx(e')$ and $D = \Ectx(e)$.
\end{proof}

\subsection{Strengthened Preservation}

We are now ready to state the strengthened preservation theorem for statements.

\begin{theorem}[Strengthened Preservation for Statements]
  \label{thm:pres-stmt}
  If all of the following hold:
  \begin{enumerate}[label=(\alph*)]
  \item \label{thm:pres-stmt-hstmt} $\tjudgS{\PrfCtxS}{s}{\Env;L',\Guards}$
  \item \label{thm:pres-stmt-hstore} $\tjudgStore{\store}$
  \item \label{thm:pres-stmt-hstep} $\stepS{s}{\store}{s'}{\store'}{\store_{sub}}$
  \end{enumerate}
  then there exist $\PrfGdS', \Env', \Guards'$ such that
  \begin{enumerate}[label=(\roman*)]
  \item \label{thm:pres-stmt-cstmt} $\tjudgS{\epsilon;\PrfGdS'}{s'}{\Env';L',\Guards'}$
  \item \label{thm:pres-stmt-cstore} $\tjudgStore{\store'}$
  \item \label{thm:pres-stmt-cenv}
  For all $x \in \dom(\Env')$ such that $\Env(x) = \type$, it must be the case that $\Env'(x) = \store_{sub}(\type)$
  \item \label{thm:pres-stmt-csub1}
  $\dom(\stmtSub) \subseteq \dom(\Env)$
  \item \label{thm:pres-stmt-csub2}
  $\dom(\stmtSub) \cap \dom(\Env') = \emptyset$
  \item \label{thm:pres-stmt-cguards} 
  There exist $A, D$ such that $\GuardChange{A}{D}{\Guards}{\Guards'}$
  \item \label{thm:pres-stmt-cstrength} $\Encode(\Env') \land \Encode(\Guards') \implies \Encode(\stmtSub(\Guards))$ is valid
  \end{enumerate}
\end{theorem}
Intuitively, Theorem~\ref{thm:pres-stmt} says that if $s$ is well-typed and can step to $s'$, then $s'$ is also well-typed with output environment $\Env'$ and guards $\Guards'$.
Clauses \ref{thm:pres-stmt-hstmt}-\ref{thm:pres-stmt-hstep} and \ref{thm:pres-stmt-cstmt}-\ref{thm:pres-stmt-cstore} are standard, except that $\Env'$ and $\Guards'$ may be different from $\Env$ and $\Guards$ with several restrictions.
The next three clauses, \ref{thm:pres-stmt-cenv}-\ref{thm:pres-stmt-csub2}, concern the environment $\Env'$.
The substituted variables $\stmtSub$ must be bound in $\Env$ (clause \ref{thm:pres-stmt-csub1}); since they are deleted, they may no longer be bound in $\Env'$ (clause \ref{thm:pres-stmt-csub2}).
Any variables that are bound in $\Env$ but remain bound in $\Env'$ must have the same type up to substitution (clause \ref{thm:pres-stmt-cenv}).

The last two clauses constrain how the output $\Guards'$ may change.
In particular, clause~\ref{thm:pres-stmt-cguards} requires that the only difference between $\Guards$ and $\Guards'$ are the predicates $D$ deleted and the predicates $A$ added, and clause~\ref{thm:pres-stmt-cstrength} forces the new guards $\Guards'$ after executing $s'$ to be stronger than the initial guards $\Guards$ (up to substitution) after executing $s$.
This notion of "precondition strengthening" allows redundant predicates to be removed without affecting well-typedness and ensures that sufficient predicates are not removed.

\begin{proof} By induction on the derivation of  \ref{thm:pres-stmt-hstep}. Note that goal~\ref{thm:pres-stmt-cstore} is trivially discharged when $\store' = \store$; goals~\ref{thm:pres-stmt-cenv}, \ref{thm:pres-stmt-csub1}, \ref{thm:pres-stmt-csub2} are trivially discharged when $\stmtSub = \epsilon$.; and goal~\ref{thm:pres-stmt-cstrength} is trivially discharged when $\stmtSub = \epsilon$ and $\Guards = \Guards'$.

By case analysis on \ref{thm:pres-stmt-hstep}:
\begin{itemize}
\item Case {\sc ES-Ctx}: Then
\begin{align}
    s &= \Sctx(e) \\
    \stepE{e&}{e'} \label{eq:prf-spres-ctx-step} \\
    \store' &= \store \\
    \store_{sub} &= \epsilon \\
    s' &= \Sctx(e')
\end{align}
By Lemma~\ref{lem:eval-sctx-arg-typed}, there exists a $\type$ such that
\begin{equation}
    \tjudg{\PrfCtxS}{\expr}{\type} \label{eq:prf-spres-ctx-e-type}
\end{equation}
By applying Theorem~\ref{thm:pres-expr} to Eq.~\eqref{eq:prf-spres-ctx-step} and Eq.~\eqref{eq:prf-spres-ctx-e-type},
\begin{equation}
    \tjudg{\PrfCtxS}{\expr'}{\type}
\end{equation}
By \ref{thm:pres-stmt-hstmt} and Lemma~\ref{lem:eval-sctx-swap-typed}, there exists $\Guards'$ such that
\begin{align}
    \tjudgS{\PrfCtxS}{\Ectx_s(e')}{\Env;\Guards'} \\
    \GuardChange{A}{D}{\Guards}{\Guards'}
\end{align}
The only remaining goal is~\ref{thm:pres-stmt-cstrength}, which follows by considering the values of $A$ and $D$ as shown in the proof of Lemma~\ref{lem:eval-sctx-swap-typed}.

\item Cases {\sc ES-AssertTrue, ES-AssumeTrue}: Follows immediately by application of \textsc{TS-Skip}.

\item Case {\sc ES-Seq1}: Then
\begin{align}
    s &= s_1;s_2 \\
    \stepS{s_1}{\store&}{s_1'}{\store'}{\store_{sub}} \\
    \store_{sub} &= (x_1 \mapsto v_1, \dots, x_n \mapsto v_n) \\
    s' &= s_1';\stmtSub(s_2)
\end{align}
By inversion on the derivation of  \ref{thm:pres-stmt-hstmt}, we see that the derivation must have used \textsc{TS-Seq}:
\begin{align}
    \tjudgS{\PrfCtxS&}{s_1}{\Env_1;L_1;\Guards_1} \label{eq:prf-spres-seq1-s1type} \\
    \tjudgS{\Env_1;L_1,\Guards_1&}{s_2}{\Env;L',\Guards} \label{eq:prf-spres-seq1-s2type}
\end{align}
By the inductive hypothesis and Eq.~\eqref{eq:prf-spres-seq1-s1type}, there exist $\PrfGdS', \Env_1', \Guards_1', A_1, D_1$ such that
\begin{align}
    \tjudgS{\epsilon;\PrfGdS'}{&s_1'}{\Env_1';L_1,\Guards_1'} \label{eq:prf-spres-seq1-s1p} \\
    \tjudgStore{&\store'} \\
    \GuardChange{A_1}{D_1}{\Guards_1&}{\Guards_1'} \label{eq:prf-spres-seq1-gchange} \\
    \dom(\stmtSub) &\subseteq \Env_1 \label{eq:prf-spres-seq1-subenv1} \\
    \dom(\stmtSub) &\cap \Env_1' = \emptyset \\
    \Encode(\Env_1') \land \Encode(\Guards_1') &\implies \Encode(\stmtSub(\Guards_1)) \text{ valid} \label{eq:prf-spres-seq1-strength}
\end{align}
discharging goal~\ref{thm:pres-stmt-cstore}.

By Eq.~\eqref{eq:prf-spres-seq1-gchange}, there exists $\Guards_1''$ such that
\begin{align}
    \Guards_1 &= D_1, \Guards_1'' \\
    \Guards_1' &= A_1, \Guards_1''
\end{align}
By Eq.~\eqref{eq:prf-spres-seq1-subenv1} and Lemma~\ref{lem:persistent-bind}, there exist $\Env'', \Env_1''$, $\Guards''$ such that
\begin{align}
    \Env_1 &= x_1: \type_1, \dots, x_n: \type_n, \Env_1'' \\
    \Env &= x_1: \type_1, \dots, x_n: \type_n, \Env_1'', \Env''
    \\
    \Guards &= \Guards_1, \Guards'' \label{eq:prf-spres-seq1-split-guards}
\end{align}
By Eq.~\eqref{eq:prf-spres-seq1-s2type} and Lemma~\ref{lem:subst-stmt},
\begin{align}
    \tjudgS{\stmtSub(\Env_1');L_1,\stmtSub(\Guards_1)&}{\stmtSub(s_2)}{\stmtSub(\Env);L',\stmtSub(\Guards)}
\end{align}
By Eq.~\eqref{eq:prf-spres-seq1-strength}, Lemma~\ref{lem:weaken-stmt2}, and Eq.~\eqref{eq:prf-spres-seq1-split-guards},
\begin{align}
    \tjudgS{\stmtSub(\Env_1');L_1,\Guards_1'&}{\stmtSub(s_2)}{\stmtSub(\Env);L',\Guards_1',\stmtSub(\Guards'')} \label{eq:prf-spres-seq1-s2s}
\end{align}

Use $\PrfGdS'$ as before, and set $\Env' = \stmtSub(\Env)$ and $\Guards' = \Guards_1', \stmtSub(\Guards'')$; this discharges goals~\ref{thm:pres-stmt-csub1} and \ref{thm:pres-stmt-csub2}.
Furthermore, goal~\ref{thm:pres-stmt-cstrength} follows immediately from Eqs.~\eqref{eq:prf-spres-seq1-split-guards} and \eqref{eq:prf-spres-seq1-strength}.

Note that $\stmtSub(\Env_1') = \Env_1$ since $\dom(\stmtSub) \cap \dom(\Env_1) = \emptyset$.
Thus, applying {\sc TS-Seq} to Eq.~\eqref{eq:prf-spres-seq1-s1p} and Eq.~\eqref{eq:prf-spres-seq1-s2s} discharges goal~\ref{thm:pres-stmt-cstmt}.

Finally, setting $A = \Guards'$ and $D = \Guards$ discharges goal~\ref{thm:pres-stmt-cguards}.

\item Case {\sc ES-Seq2}: Then $s = \sskip; s_2$ and $s' = s_2$, so the result follows by inversion of the derivation of \ref{thm:pres-stmt-hstmt}.

\item Case {\sc ES-Seq3}: Then
\begin{align}
    s &= \sassume \efalse; s_2 \\
    s' &= \sassume \efalse \\
    \store' &= \store \\
    \store_{sub} &= \epsilon
\end{align}
Setting $\PrfGdS' = \PrfGdS$, $A = \efalse$, $D = \Guards$, $\Guards' = \efalse$, and $\Env = \epsilon$ and applying {\sc TS-Assume} discharges the remaining goals.

\item Case {\sc ES-Let}: Then
\begin{align}
    s &= \slet{x}{\type}{v} \\
    s' &= \sskip \\
    \store' &= \store \\
    \store_{sub} &= (x \mapsto v)
\end{align}
discharging goal~\ref{thm:pres-stmt-cstore}.
By inversion on the derivation of~\ref{thm:pres-stmt-hstmt},
\begin{align}
    \Env &= x : \type \\
    \Guards &= \epsilon \\
    L' &= \PrfGdS
\end{align}
discharging goal~\ref{thm:pres-stmt-csub1}.
Setting $\PrfGdS' = \PrfGdS$, $\Guards' = \epsilon$, and $\Env' = \epsilon$ and applying \textsc{TS-Skip} discharges the remaining goals.

\item Case {\sc ES-Fetch}: Similar to \textsc{ES-Let}.

\item Case {\sc ES-Commit}: Then
\begin{align}
  s &= \scommit{v_1 \sto x_1, \dots, v_n \sto x_n} \\
  s' &= \sskip \\
  \dom(\store) &= \{x_1, \dots, x_n\} \\
  \store_{sub} &= \epsilon \\
  \store' &= (x_1 \mapsto v_1, \dots, x_n \mapsto v_n)
\end{align}
By inversion on the derivation of \ref{thm:pres-stmt-hstmt}, we see that the derivation must have used \textsc{TS-Commit}:
\begin{align}
  \defns(x_i) &= \type_i' & \text{for $i = 1..n$} \\
  \tjudg{\PrfCtxS&}{v_i}{\type_i} & \text{for $i = 1..n$} \\
  \subty{\PrfCtxS&}{\type_i}{\substp{\type_i'}{\subst{x_1}{v_1}, \dots, \subst{x_i}{\nu}, \dots, \subst{x_n}{v_n}}} & \text{for $i = 1..n$} \\
  \Env &= \epsilon \\
  \Guards &= \epsilon \\
  \PrfGdS &= \locked \\
  L' &= \unlocked
\end{align}
Setting $\PrfGdS' = \unlocked$, $\Guards' = \epsilon$, and $\Env' = \epsilon$ and applying {\sc TS-Skip} discharges most goals.
To show \ref{thm:pres-stmt-cstore}, observe that the free variables of the refinements of $\type_1', \dots \type_n'$ are a subset of $\{x_1, \dots, x_n\}$. Thus, we can obtain the required result by inverting the derivation of $\tjudgStore{\store}$ and applying \textsc{T-Store}.

\item Case {\sc ES-Call}:
Then
\begin{align}
    s &= \scall{x:\type_r}{f}{v_1, \dots, v_n} \\
    \defnsE(f) &= \declFun{f}{x_1:\ \type_1, \dots, x_n:\ \type_n}{\type_r}{s_f}{e} \\
    \stmtSub' &= (\subst{x_1}{v_1}, \dots, \subst{x_n}{v_n}) \\
    s' &= \stmtSub'(s_f);\slet{x}{\type}{e} \\
    \store' &= \store \\
    \store_{sub} &= \epsilon
\end{align}
discharging goals~\ref{thm:pres-stmt-cstore}, \ref{thm:pres-stmt-csub1}, \ref{thm:pres-stmt-csub2}.
By inversion on the derivation of $\tjudgS{\PrfCtxS}{s}{\Env;\PrfGdS',\Guards}$, we see that the derivation must have used \textsc{TS-Call}:
\begin{align}
  \Env &= x : \tau_r[x_1' \mapsto v_1, \dots, x_n' \mapsto v_n] \\
  \PrfGdS &= \PrfGdS' = \unlocked \\
  \Guards &= \epsilon \\
  \defns(f) &= ((x_1' \mapsto \type_1', \dotsi, x_n' \mapsto \type_n'), \tau_r)
  \\
  \tjudg{\PrfCtxS&}{v_i}{\type_i} & \text{for $i = 1..n$}
  \\
  \subty{\PrfCtxS&}{\type_i}{\type_i'[x_1' \mapsto v_1, \dots, x_i' \mapsto \nu, \dots, x_n' \mapsto v_n]} & \text{for $i = 1..n$}
\end{align}
discharging goal~\ref{thm:pres-stmt-cstrength}.
By inversion on $\tjudgGlobals{\defnsE}$, we see that $f$ is well-typed using \textsc{T-FunDecl}:
\begin{align}
    \tjudgS{x_1:\type_1,\dots,x_n:\type_n;\unlocked&}{s_f}{\Env_f;\unlocked,\Guards_f} \label{eq:prf-pres-call-sf} \\
    \tjudg{\Env_f;\Guards_f&}{e}{\type_r'} \\
    \subty{\Env_f;\Guards_f&}{\type_r'}{\type_r}
\end{align}
By Lemma~\ref{lem:subst-stmt} and Eq.~\eqref{eq:prf-pres-call-sf},
\begin{align}
    \tjudgS{\epsilon;\unlocked&}{\stmtSub'(s_f)}{\stmtSub'(\Env_f);\unlocked,\stmtSub'(\Guards_f)}
    \label{eq:prf-pres-call-sf-type}
    \\
    \tjudg{\stmtSub'(\Env_f);\stmtSub'(\Guards_f)&}{\stmtSub'(e)}{\type_r'}
\end{align}
Applying {\sc TS-Let} yields
\begin{align}
    \tjudgS{\stmtSub'(\Env_f);\unlocked,\stmtSub'(\Guards_f)&}{\slet{x}{\type_r}{e}}{x:\type_r,\stmtSub'(\Env_f);\unlocked,\stmtSub'(\Guards_f)}
    \label{eq:prf-pres-call-lt-type}
\end{align}
Applying {\sc TS-Seq} to Eq.~\eqref{eq:prf-pres-call-sf-type} and Eq.~\eqref{eq:prf-pres-call-lt-type} and noting that $\stmtSub'(s_f) = \substp{s_f}{\subst{x_1}{v_1}, \dots, \subst{x_n}{v_n}}$ yields
\begin{align}
    \tjudgS{\epsilon;\unlocked&}{s'}{x:\type_r,\stmtSub'(\Env_f);\unlocked,\stmtSub'(\Guards_f)}
\end{align}
Setting $\Env' = \stmtSub'(\Env_f)$ and $\Guards' = \stmtSub'(\Guards_f)$ discharges goals~\ref{thm:pres-stmt-cstmt}, \ref{thm:pres-stmt-cenv}. 
Finally, setting $A = \Guards'$ and $D = \epsilon$ discharges goal~\ref{thm:pres-stmt-cguards}.

\item Case {\sc ES-IfTrue}: Then
\begin{align}
    s &= \sif{\etrue}{s_1}{s_2}{j} \\
    j &= x_1 : \type_1 = \phi(x_{11}, x_{12}), \dots, x_n : \type_n = \phi(x_{n1}, x_{n2}) \\
    s_j &= \slet{x_1}{\type_1}{x_{11}};\dots;\slet{x_n}{\type_n}{x_{n1}} \label{eq:prf-pres-iftrue-sj} \\
    s' &= s_1; s_j \\
    \store' &= \store \\
    \stmtSub &= \epsilon
\end{align}
discharging goals \ref{thm:pres-stmt-cstore}, \ref{thm:pres-stmt-cenv}, \ref{thm:pres-stmt-csub1}, \ref{thm:pres-stmt-csub2}.
By inversion on the derivation of  \ref{thm:pres-stmt-hstmt}, we must have used {\sc TS-If}:
\begin{align}
    \tjudgS{\epsilon;\PrfGdS,\etrue&}{s_1}{\Env_1;\PrfGdS',\Guards_1} \label{eq:prf-pres-iftrue-s1-true} \\
    \tjudgS{\epsilon;\PrfGdS,\efalse&}{s_2}{\Env_2;\PrfGdS',\Guards_2} \\
    \tjudgS{\epsilon;\PrfGdS;(\Env_1;\PrfGdS',\Guards_1);(\Env_2;\PrfGdS',\Guards_2)&}{j}{\Env} \label{eq:prf-pres-iftrue-j} \\
    \Guards &= \epsilon
\end{align}
By inversion on the derivation of Eq.~\eqref{eq:prf-pres-iftrue-j},
\begin{align}
    \Env &= x_1: \type_1, \dots, x_n: \type_n \\
\end{align}
Setting $\Env' = x_1: \type_1, \dots, x_n : \type_n, \Env_1$ discharges goal~\ref{thm:pres-stmt-cenv}.
Then, applying {\sc TS-Let} and {\sc TS-Seq} to Eq.~\eqref{eq:prf-pres-iftrue-sj} multiple times yields
\begin{align}
    \tjudgS{\Env_1;\PrfGdS', \Guards_1&}{s_j}{\Env';\PrfGdS',\Guards_1}
    \label{eq:prf-pres-iftrue-sj-type}
\end{align}
Applying {\sc TS-Seq} to Eqs.~\eqref{eq:prf-pres-iftrue-s1-true} and \eqref{eq:prf-pres-iftrue-sj-type} yields
\begin{align}
    \tjudgS{\epsilon;\PrfGdS, \etrue&}{s_1;s_j}{\Env';\PrfGdS',\Guards_1}
\end{align}
By Lemma~\ref{lem:persistent-bind} and Lemma~\ref{lem:contract-true-stmt}, there exists $\Guards'$ such that
\begin{align}
    \tjudgS{\PrfCtxS&}{s_1;s_j}{\Env';\PrfGdS',\Guards'}
\end{align}
discharging goal~\ref{thm:pres-stmt-cstmt}.
Finally, setting $A = \Guards'$ and $D = \epsilon$ discharges goals~\ref{thm:pres-stmt-cguards}, \ref{thm:pres-stmt-cstrength}.

\item Case {\sc ES-IfFalse}: Similar.
\end{itemize}
\end{proof}

\fi

\end{document}